\documentclass[11pt]{article}
\usepackage{setspace}

\pdfoutput=1
\usepackage[utf8]{inputenc}
\usepackage[icelandic, english]{babel}
\usepackage{t1enc}
\usepackage{graphicx}
\usepackage[intoc]{nomencl}
\usepackage{enumerate,color}
\usepackage{url}
\usepackage{amsmath}
\usepackage{amsfonts}
\usepackage{thmtools}
\usepackage{amssymb}
\usepackage[nottoc]{tocbibind}
\usepackage[sf,normalsize]{subfigure}
\usepackage[format=plain,labelformat=simple,labelsep=colon]{caption}
\usepackage{placeins}
\usepackage{tabularx}
\usepackage[T1]{fontenc}
\usepackage{algorithm}
\usepackage[noend]{algpseudocode}
\usepackage{natbib}
\usepackage{booktabs}
\usepackage[table,dvipsnames]{xcolor}
\usepackage{aps-bibstyle}   

\newcommand{\fat}[1]{\boldsymbol{#1}}
\newcommand{\trp}{^\mathsf{T}}

\newcounter{lastnote}

\title{Approximate Bayesian inference for analysis of spatio-temporal flood frequency data} 

\author
{\'Arni V. Johannesson$^{1}$, Stefan Siegert$^{2}$, Rapha\"el Huser$^{3\ast}$, \\
Haakon Bakka$^{4}$, Birgir Hrafnkelsson,$^{1}$ \\
\normalsize{$^{1}$University of Iceland,}\\
\normalsize{$^{2}$University of Exeter}\\
\normalsize{$^{3}$King Abdullah University of Science and Technology}\\
\normalsize{$^{4}$University of Oslo}\\
}

\date{}

\begin{document} 




\baselineskip12pt


\maketitle 


\begin{abstract}
Extreme floods cause casualties, and widespread damage to property and vital civil infrastructure. Predictions of extreme floods within gauged and ungauged catchments is crucial to mitigate these disasters. In this paper, a Bayesian framework is proposed for predicting extreme floods using the generalized extreme-value (GEV) distribution. A major methodological challenge is to find a suitable parametrization for the GEV distribution when multiple covariates and/or latent spatial effects are involved and a time trend is present. Other challenges involve balancing model complexity and parsimony using an appropriate model selection procedure, and making inference based on a reliable and computationally efficient approach. We here propose a latent Gaussian modeling framework with a novel multivariate link function designed to separate the interpretation of the parameters at the latent level and to avoid unreasonable estimates of the shape and time trend parameters. Structured additive regression models, which include catchment descriptors as covariates and spatially correlated model components, are proposed for the four parameters at the latent level. To achieve computational efficiency with large datasets and richly parametrized models, we exploit a highly accurate and fast approximate Bayesian inference approach, which can also be used to efficiently select models separately for each of the four regression models at the latent level. We applied our proposed methodology to annual peak river flow data from 554 catchments across the United Kingdom. The framework performed well in terms of flood predictions for both ungauged catchments and future observations at gauged catchments. The results show that the spatial model components for the transformed location and scale parameters, as well as the time trend, are all important and none of these should be ignored. Posterior estimates of the time trend parameters correspond to an average increase of about $1.5$\% per decade with range $0.1$\% to $2.8$\% and reveal a spatial structure across the United Kingdom. When the interest lies in estimating return levels for spatial aggregates, we further develop a novel copula-based post-processing approach of posterior predictive samples, in order to mitigate the effect of the conditional independence assumption at the data level, and we demonstrate that our approach indeed provides accurate results.
\end{abstract}

\section{Introduction}

Flood predictions are of great importance for the design, management, operation, and protection of vital infrastructure. To aid flood predictions, gauging stations have been built to observe and monitor streams, wells, lakes, canals and reservoirs. Flood predictions are typically based on analyses of past observed extremes, and they usually rely on fitting statistical models for extreme floods within gauged catchments. Predictions for ungauged catchments are made possible by using appropriate covariates whose coefficients are learned from gauged catchments, and sometimes also by exploiting spatial correlation across catchments using suitable geostatistical models. 


One of the first approaches in flood frequency analysis was the index flood method introduced in \cite{dalrymple1960flood}. The index flood method is a statistical method that was designed to handle cases where little or no site-specific information is available, by borrowing strength from similar catchments. The index flood method consists of two steps: Regionalization, which identifies geographically and climatologically similar regions; and specification of a regional standardized flood frequency curve for the $T$-year return level. Various developments have been made to improve the index flood method \citep{hosking1985estimation, grehy1996presentation}, and various Bayesian methods have been introduced \citep{cunnane1974bayesian, rosbjerg1995uncertainty, kuczera1999comprehensive, martins2000generalized}. 
In the United Kingdom (UK), the analysis of floods is commonly tackled with the pooling group method \citep{robson1999flood, kjeldsen2009formal}, which is a combination of (i) the index flood method based on the use of $L$-moments 
\citep{hosking2005regional}; and (ii) the region of influence (ROI) approach 
\citep{burn1990evaluation}. 
Further related statistical approaches for the modeling, prediction, and uncertainty assessment of annual peak flow data within gauged and ungauged catchments were proposed by, e.g., \citet{kjeldsen2006prediction,kjeldsen2009exploratory}, \citet{kjeldsen2010modelling}, and \citet{kjeldsen2017use}. More recently, \citet{thorarinsdottir2018bayesian} analyzed annual maximum floods from $203$ streamflow stations in Norway based on a Bayesian latent Gaussian model (LGM) with a generalized extreme-value (GEV) distribution at the data level, which is motivated by Extreme-Value Theory (see, e.g., \citet{Davison.etal:2012} and \citet{Davison.Huser:2015}). They used a multivariate link function specified by identity links for the location and shape parameters and a logarithmic link for the scale parameter. Their model is fairly simple in the sense that it does not contain any spatially correlated random effects (a priori) and ignores the presence of a time trend, and inference is performed by drawing samples from the posterior distribution using a Markov chain Monte Carlo (MCMC) algorithm. Bayesian models for extremes with spatially correlated random effects date back to \cite{casson1999}. More recently, models of this sort were also considered by \citet{Huerta2007}, \citet{Davison.etal:2012}, \citet{geirsson2015computationally} and \citet{Dyrrdal.etal:2015} in studies of extreme ozone concentrations and/or precipitation but the datasets in these papers were limited to $19$, $36$, $40$ and $59$ stations only, respectively; see also \cite{hrafnkelsson2012spatial}, who used a similar approach for modeling annual minimum and maximum temperatures observed at $72$ stations in Iceland. \citet{Sang2009} modeled extreme precipitation observed on a large regular grid with $1078$ grid cells by taking advantage of conditional autoregressive (CAR) priors for the location and scale parameters, along with an additional temporal trend term for the location parameter, and assuming a constant shape parameter. \citet{Sang2010} then relaxed the conditional independence assumption by using a Gaussian dependence structure at the data level. In the same vein, \citet{Cooley.Sain:2010} and \citet{Jalbert.etal:2017} modeled extreme precipitation data arising from climate model outputs over large grids, by exploiting their gridded structure using intrinsic Gaussian Markov random field (IGMRF) priors at the latent level of the LGM.


In this paper we propose a novel statistical model for annual peak flow data from $554$ catchments across the UK that efficiently exploits spatial information about the catchments. Similarly to \citet{Huerta2007}, \citet{Sang2009,Sang2010}, \citet{Davison.etal:2012}, \citet{hrafnkelsson2012spatial}, \citet{Dyrrdal.etal:2015}, \citet{geirsson2015computationally}, \citet{Jalbert.etal:2017} and \citet{thorarinsdottir2018bayesian}, our proposed model is an LGM that assumes a GEV distribution at the data level with site-specific location (here, with intercept and time trend), scale and shape parameters, but its detailed hierarchical structure and our inference procedure have crucial differences that make it more attractive and amenable to complex and large datasets. Moreover, contrary to \citet{Sang2009,Sang2010}, \citet{Cooley.Sain:2010} and \citet{Jalbert.etal:2017}, the data in our application are not regularly located, which makes it more challenging from a computational perspective. At the latent level, we follow \citet{rigby2005generalized} and use a generalized additive specification for the transformed location intercept, scale, shape {and trend} parameters of the data density.   
Our proposed model at the latent level contains catchment descriptors as covariates, e.g., catchment area and average annual rainfall, along with spatial model components for the transformed location intercept and scale parameters, which are specified as approximate solutions to stochastic partial differential equations (SPDEs, see \cite{lindgren2011explicit}). These latent SPDE-based model components can capture complex spatially structured residuals more flexibly than the CAR priors of \citet{Sang2009,Sang2010}, and also yield sparse precision matrices, which accelerates computations with large spatial datasets. Because the four parameters of the data level (i.e., the GEV location intercept and time trend, scale and shape parameters) are transformed to the latent level, the proposed model is an LGM with a multivariate link function, referred to here as an extended LGM \citep{geirsson2020mcmc,hrafnkelsson2020approx}.

{The multivariate link function consists of four transformations.
The logarithmic transformation is applied to the GEV location parameter, $\mu$. 
The other three transformations have not been presented before in this context, and can thus be considered novel. The shape parameter, $\xi$, and the time trend parameter, $\Delta$, are assumed to take values only within specified intervals, and their transformations reflect that. The specified ranges for $\xi$ and $\Delta$ are selected to eliminate unnecessarily small or large parameter values where the aim is twofold: (i) reducing parameter uncertainty; and (ii) stabilizing site-wise likelihood fits.}
Our proposed link function keeps an intuitive interpretation and desirable properties for the shape parameter, while producing reliable estimates when several covariates are involved. Moreover, to prevent overfitting when multiple spatial random effects are involved, we specify relatively informative shrinkage priors for the hyperparameters, in order to penalize complex models that depart too much from a simpler counterpart \citep{simpson2017penalising}.

In the frequentist framework, the \texttt{gevlss} function of the \texttt{R} package \texttt{mgcv} may be used to fit generalized extreme value location, scale and shape models. More recently, a fast frequentist method based on restricted maximum likelihood was developed for models similar to the model proposed in this paper \citep{youngman2019gam}. Alternatively, in the Bayesian framework, MCMC methods may be exploited. However, instead of inferring our proposed Bayesian extended LGM with an ``exact'' MCMC sampler as in \cite{geirsson2020mcmc}, we here exploit {an accurate} Gaussian approximation to the posterior density adapted from the general methodology recently proposed by \citet{hrafnkelsson2020approx}, in order to better handle the high dimensionality of the data and to improve mixing and convergence of the MCMC chains.
{It is the first time that this methodology, called ``Max-and-Smooth'', is applied and carefully validated in the context of large-scale, heavy-tailed, non-stationary, spatially referenced data.}

This fast Gaussian-based approximation of the posterior density is also used to break the selection of catchment descriptors down such that the descriptors for one of the four transformed parameters at the latent level are {selected} separately of the other three. This separation makes it possible to apply the integrated nested Laplace approximation (INLA, \citealp{rue2009approximate}) to each of the transformed parameters and a $10$-fold cross-validation experiment for the selection of catchment descriptors becomes feasible. {We consider this to be a novel aspect of our analysis.} 
Furthermore, the computation of quantiles {(i.e., marginal return levels)} is straightforward and their prediction at ungauged sites can be made efficiently {and accurately} through the posterior predictive distributions using site-specific covariates. 

{The spatial and temporal predictive performances of the proposed latent Gaussian model are evaluated in a detailed cross-validation study by comparing it to other natural alternative approaches based on the GEV distribution, namely three frequentist models (constant model; site-wise model; response surface model), and three simplified version of the proposed LGM (without trend; without spatial components; without covariates). These new results provide strong support in favor of our approach and clearly indicate that our proposed model performs well and provides significant improvements over other state-of-the-art approaches.} 

{Whilst our main objective in this work is to estimate marginal return levels of river flow at each individual site, it may sometimes be required for planning or regional flood risk assessment to estimate return levels for spatial aggregates. In this case, both marginal distributions and the spatial dependence structure play a key role. Therefore, while the conditional independence assumption that underpins our proposed model at the data level has a minor effect on marginal return level estimates, it does however strongly affect the estimation of spatial return levels. To circumvent this limitation, we here further develop a novel copula-based method, which consists in post-processing posterior predictive samples from our model by simply matching their ranks with those of the observed data, in such a way to accurately ``correct'' the bias and variance in estimated spatial return levels.}

The paper is organized as follows. In Section \ref{ch:data}, we present the data and covariates used in our statistical analysis. We then describe our specific flood frequency model in Section \ref{ch:mdlSpec}. In Section \ref{ch:inference}, we present the exact posterior density and its approximation, and we also introduce the method for selecting covariates for our flood frequency model. {The results of the model selection, the posterior inference, a detailed cross-validation study to compare our model to natural alternatives, as well as our proposed post-processing approach of posterior predictive samples for estimating spatial return levels, are all} presented and discussed in Section \ref{ch:results}. The paper concludes with a discussion and perspectives for future research in Section \ref{ch:discussion}.

\section{Data and exploratory analysis}
\label{ch:data}

\begin{figure}[t!]
\centering
\includegraphics[width=0.6\textwidth]{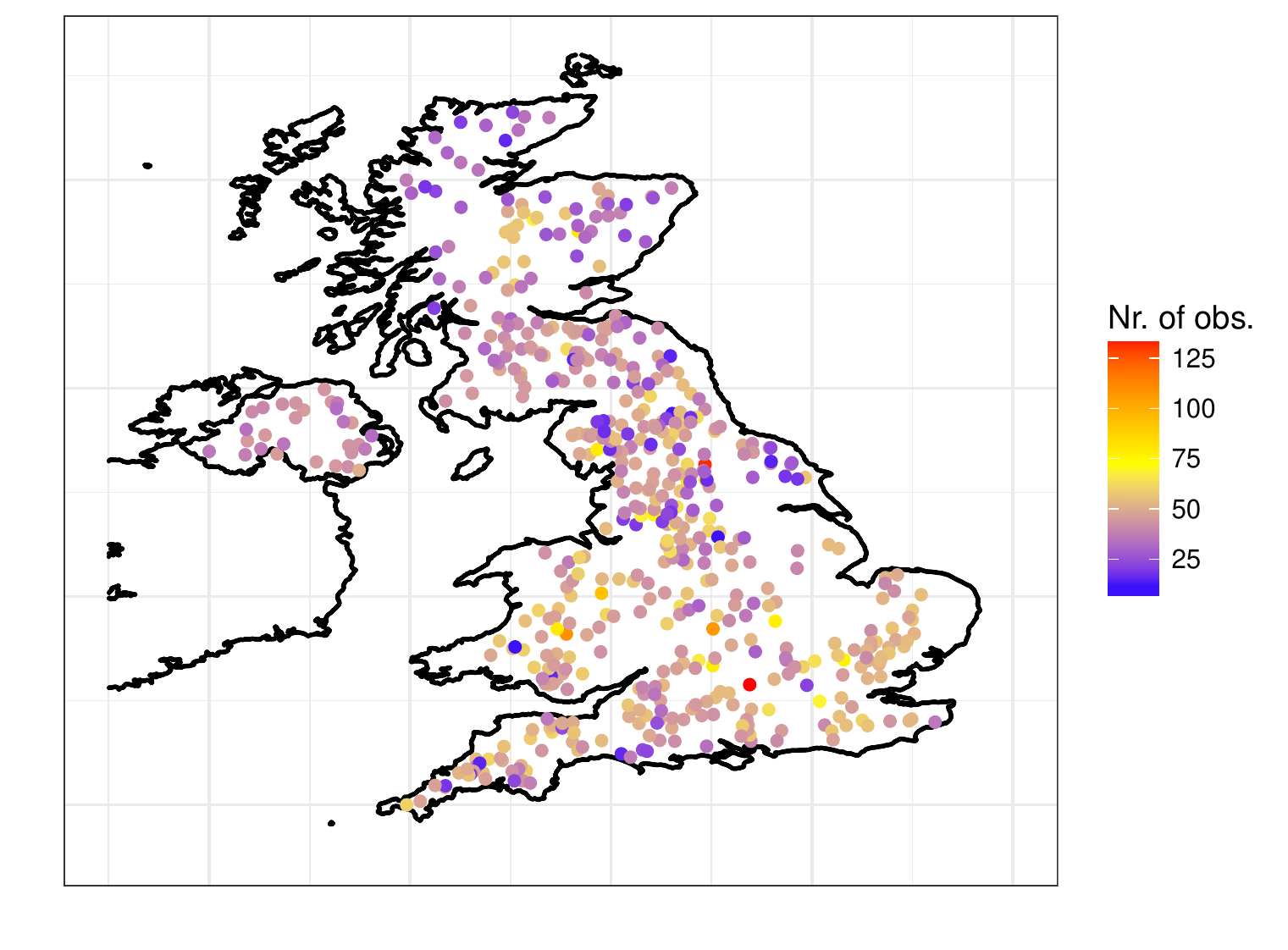}
\caption{Map of the UK with the $554$ stations used in our flood frequency analysis. The color bar indicates the number of annual maximum flow observations available at each station.} \label{fig:stationLocation}
\end{figure}

The data used in this paper come from the UK National River Flow Archive which is hosted by the Centre for Ecology \& Hydrology on behalf of the Natural Environment Research Council; see \citet{nrfa}. The data consist of annual maxima from daily river flow observations measured at $554$ gauging stations located across the UK. The observations date back to 1851 but most of the observations were obtained from 1970 to 2013. A large fraction of the observational sites have between $50$ and $80$ annual maxima available. The dataset actually contains 958 gauging stations, but only stations that are deemed suitable for pooling and index flood calculations due to their homogeneous quality are used in our analysis. These stations are explicitly flagged as such in the original dataset. 
Figure \ref{fig:stationLocation} shows the location of the $554$ stations along with the number of annual maxima available at each station. 

The \citet{anderson1954test} goodness-of-fit test, as presented by \citet{stephens1974edf}, was applied across the observational sites. It revealed that the GEV distribution with a time trend is an appropriate model for the annual maxima data; see {Section 2.2 of} the Supplementary Material for more details. This is no surprise, because the GEV distribution is the only possible limiting distribution for renormalized block maxima. The GEV distribution has been applied successfully to a wide range of problems; see, e.g., the book of \citet{Coles:2001}, the review paper by \citet{Davison.Huser:2015}, and references therein. The GEV cumulative distribution function may be expressed as
\begin{equation}
\label{eq:GEV}
\text{GEV}(y\mid \mu,\sigma,\xi)=\left\{\begin{array}{ll}
\exp\left[-\left\{1+\xi(y-\mu)/\sigma\right\}_+^{-1/\xi}\right], & \xi\neq0;\\
\exp\left[-\exp\left\{-(y-\mu)/\sigma\right\}\right], & \xi=0,
\end{array}\right.
\end{equation}
where $\mu\in\mathbb R$, $\sigma>0$ and $\xi\in\mathbb R$ are location, scale and shape parameters, respectively, and $a_+=\max(a,0)$. The support of the GEV distribution is $[-\infty,\mu-\sigma/\xi]$, $[-\infty,+\infty]$, and $[\mu-\sigma/\xi,+\infty]$ when $\xi<0$, $\xi=0$, and $\xi>0$, respectively, and {$\xi$ determines} the tail heaviness, with a short, light and heavy upper tail when $\xi<0$, $\xi=0$, and $\xi>0$, respectively.

In order to propose transformations for the GEV parameters $\mu$, $\sigma$ and $\xi$, we first explore their relationships as suggested by the data. To investigate this, we fit the GEV distribution with a linear time trend in $\mu$ to all stations separately by (generalized) maximum likelihood (ML). For completeness, we report these preliminary site-wise ML estimates in {Section 2.1 of} the Supplementary Material. The estimated location parameters are all positive and highly right-skewed, and a time trend is often present. Thus, we model the location parameter at location $i$ and time $t$ as $\mu_{it}=\mu_i\{1+\Delta_i(t-t_0)\}$ where $\mu_i$ is a site-specific location intercept at time $t=t_0$, the reference year is set to $t_0=1975$, and $\Delta_i$ is a site-specific time trend parameter. We use the logarithmic transformation $\psi_i=\log(\mu_i)$ and a scaled logit transformation to transform $\Delta_i$ to $\gamma_i$ such that $\Delta_i$ lies in the interval $(-0.008,0.008)$ for all stations $i$. {This corresponds to $8$\% increase or decrease in $\mu_i$ per decade.}
{Most of the UK flood datasets are observed during the years from 1950 to 2013. A trend line with $\Delta=0.008$ means that $\mu$ will change by a factor of $1.63$ over the period 1950--2013. The same comparison with $\Delta=-0.008$ gives the factor $0.58$. These two extreme scenarios correspond to a substantial increase and decrease over time for a natural process that is driven by meteorological variables. Thus, the interval $[-0.008,0.008]$ contains plausible values for the trend parameter $\Delta$.} Our Bayesian model is fully specified below in Section~\ref{ch:mdlSpec}. Moreover, the station-wise ML estimates suggest there is a clear positive linear relationship between the logarithms of the location and the scale parameters, see {Section 2.1 of} the Supplementary Material for more details. 
Thus, we propose using the transformation $\tau_i = \log(\sigma_i/\mu_i)$. When modeling $\tau_i$ with covariates and a spatial effect, we can thus capture the variability in the scale parameter that is not explained by the location parameter. {In Section \ref{ch:mdlSpec}, criteria for the shape parameter suggest that it should lie in the interval $(-0.5,0.5)$, and we thus propose a suitable function that maps site-specific shape parameter values, $\xi_i$, from the interval $(-0.5,0.5)$ to transformed shape parameter values, $\phi_i$, which are defined on the whole real line.}

The annual peak flow time series contain spatial information, and it turns out that using this information is beneficial for accurate flood predictions. The spatial features in the location, scale, shape and time trend of the data can be diagnosed by considering the variograms \citep{cressie1993statistics} of the residuals of four basic linear models
; see {Section 2.1 of} the Supplementary Material for further details. These variograms {clearly indicate} that the residuals from the linear models for the ML estimates of $\psi_i$ and $\tau_i$ are spatially correlated while the variograms of the residuals corresponding to the ML estimates of $\phi_i$ and $\gamma_i$ provide a weak indication of $\phi_i$ and $\gamma_i$ being spatially correlated. To determine which of the four parameters need spatial model components, a rigorous model selection procedure is {done; see Sections \ref{ch:mdlSel} and \ref{ch:results_mod_sel}.}

Catchment descriptors have previously been used to model extreme floods in the UK \citep[e.g.,][]{kjeldsen2010modelling, kjeldsen2017use}. In Table \ref{tab:covariateDesc}, we present all the catchment descriptors considered in {our} analysis. We selected a suitable transformation for each catchment descriptor on the basis of to their empirical {distribution} and their relationship with the ML estimates; see the caption of Table~\ref{tab:covariateDesc}. 
Most of the catchment descriptors were log-transformed. For more details on the catchment descriptors, see \cite{robson1999flood}.

The exploratory analysis briefly {outlined} above motivates the specification of our full Bayesian model, which is presented in the next section.

\begin{table}[t!]
\rowcolors{2}{gray!6}{white}
\caption{\label{tab:covariateDesc}Description of the catchment descriptors (covariates). The logarithmic function is used to transform these catchment descriptors except for BFIHOST, URBEXT and ASPBAR, which are transformed as $\textrm{BFIHOST}^2$, $\log(\textrm{URBEXT}+1)$ and $\textrm{ASPBAR}/100$, respectively.}
\centering
\fontsize{9.5}{11.5}\selectfont
\begin{tabular}[t]{>{\bfseries}p{6em}>{\raggedright\arraybackslash}p{19em}rrr}
\hiderowcolors
\toprule
\multicolumn{2}{c}{ } & \multicolumn{3}{c}{Stats.} \\
\cmidrule(l{2pt}r{2pt}){3-5}
Catchment descriptor & Description & Min & Mean & Max\\
\midrule
\showrowcolors
AREA & Catchment drainage area is the total area, in km$^2$, where the river collects water [km$^2$]. & 1.63 & 440.65 & 9930.80\\
SAAR & Average annual rainfall [mm]. & 559.00 & 1114.12 & 2913.00\\
FARL & The Flood Attenuation by Reservoirs and Lakes index quantifies the degree of flood attenuation attributable to reservoirs and lakes in the catchment. & 0.64 & 0.96 & 1.00\\
BFIHOST & A measure of catchment responsiveness. & 0.17 & 0.49 & 0.97\\
FPEXT & Floodplain extent is the fraction of the total area of the catchment that is covered by the 100-year event. & 0.00 & 0.06 & 0.30\\
URBEXT & Urban extent is an index that depicts the extent of urban/suburban land cover in the catchment area. & 0.00 & 0.03 & 0.59\\
DPLBAR & Is the mean of the distances between each grid node and the catchment outlet [km]. & 1.30 & 23.05 & 139.87\\
DPSBAR & Mean drainage path slope provides an index of steepness in the catchment, [m/km]. & 13.30 & 100.12 & 441.80\\
LDP & Longest drainage path [km]. & 2.21 & 42.94 & 286.84\\
SPRHOST & Standard percentage runoff associated with each HOST soil class. & 4.85 & 37.20 & 59.85\\
ASPBAR & The mean aspect or orientation of a catchment in degrees. & 1.00 & 151.95 & 358.00\\
ALTBAR & The mean elevation of a catchment area [m]. & 25.00 & 220.78 & 682.00\\
ASPVAR & The aspect variance in the catchment in degrees. & 0.02 & 0.17 & 0.59\\
PROPWET & An index for the proportion of time the soil in the catchment area is wet. & 0.23 & 0.48 & 0.85\\
\bottomrule
\end{tabular}
\rowcolors{2}{white}{white}
\end{table}

\section{Model specification}
\label{ch:mdlSpec}
Here we propose a novel Bayesian extended latent Gaussian model for the annual maximum flow data, where the latent level is composed of four regression models for the transformed location intercept, scale, shape and time trend parameters of the GEV density. The models for the transformed location intercept and scale parameters also contain spatial model components. 
We adopt a Bayesian framework to infer the model parameters. The Bayesian approach naturally handles simultaneous inference of multiple parameters, and  allows us to incorporate latent random effects and prior information.

Let $y_{it}$ be the annual maximum flow at station $i$ in year $t$. 
The data from site $i$ are assumed to be independent across years and to follow the GEV distribution with time-dependent location parameter $\mu_{it}$, and time-constant scale parameter $\sigma_i$ and shape parameter $\xi_i$, that is,
$$
y_{it}\sim \text{GEV}(\mu_{it},\sigma_i, \xi_i),\quad \mu_{it},\sigma_i>0,\quad \xi_i\in(-0.5,0.5),$$ where for all $y_{it}$, $1+\xi_i (y_{it}-\mu_{it})/\sigma_i>0$, $i\in \{1,\ldots,J\}$, $J$ is the number of sites, and $t\in \mathcal{A}_i$ where the set $\mathcal{A}_i$ indexes the years for which observations are available at site $i$. The location parameter $\mu_{it}$ is defined as  
$$
\mu_{it} = \mu_i\{1 + \Delta_i(t-t_0)\},
$$
where $\mu_i$ and $\Delta_i$ are the location intercept and time trend parameters for site $i$,
and $t_0 = 1975$ is a reference year. 
Note that the domains of each parameter are restricted according to the findings of our exploratory analysis conducted in Section~\ref{ch:data}. Moreover, the data are assumed to be conditionally independent given the latent GEV parameters. While this conditional independence assumption may be questioned, it is standard in the literature when the primary goal is to estimate \emph{marginal} return levels like {in this paper}, and relaxing it would make the model much more complex and intricate to fit. {We here believe that our choice of informative covariates already accounts for most of the spatial variability in the data, so that the residual spatial dependence structure is relatively weak in comparison. This is indeed confirmed by our model comparison in Section~\ref{sec:predictiveperfmodelcomparison}. Nevertheless, to rigorously assess the effect of this conditional independence assumption, we report theoretical calculations in a simplified setting} in {Section 4 of} the Supplementary Material, {showing} that this model simplification mostly has a moderate effect on the parameters' estimated uncertainty in our context. {Moreover, we further show in Section~\ref{sec:spatialdependency} how posterior predictive samples from our model can be post-processed using a copula-based approach, in order to ``correct'' their spatial dependence structure, which is key when the estimation of return levels for spatial aggregates is required.}

Similar to \cite{martins2000generalized} we use a generalized likelihood function to infer the parameters of the GEV distribution.
\cite{martins2000generalized} introduced a generalized maximum likelihood estimator as an alternative estimator in the frequentist setting for the GEV parameters. The generalized maximum likelihood estimator is based on maximizing the generalized likelihood function defined as the product of the likelihood function and
a prior density.
For flood data, \cite{martins2000generalized} argued that a beta density shifted to the interval $(-0.5,0.5)$ with mean $0.10$ and standard deviation $0.122$ is a reasonable choice; see also \citet{Cooley.Sain:2010}. Here we adopt a similar approach in 
our Bayesian inference {scheme} by replacing the likelihood function with a generalized likelihood function that includes a prior density for each $\phi_i$. These individual prior densities at each site $i$ are used in addition to the joint prior density defined at the latent level for the $\phi$ parameters. We argue below that $\xi$ should lie in $(-0.5,0.5)$, which should be reasonable in most environmental applications. {In our view, the prior distribution of \citet{martins2000generalized} is slightly too informative for our particular dataset. We opt instead for a prior distribution that is tailored to capture a wider range of tail behaviors than implied by the prior distribution of \citet{martins2000generalized}. Here, in the case of our UK flood dataset, we select a prior density that is centered around zero} with most of its mass on $(-0.3,0.3)$, as we believe a priori that values of $\xi$ close to $-0.5$ (very small upper bound) and $0.5$ (very heavy tails) are not likely. We therefore use a {symmetric} beta prior density shifted to the interval $(-0.5,0.5)$ for each $\xi_i$ with parameters $\alpha=4$ and $\beta=4$, i.e., with mean zero and standard deviation $0.167$. This beta prior density for each $\xi_i$ is then transformed to a prior density for each $\phi_i$; see {Section 1.2 of} the Supplementary Material. {This transformation is specified below.} 

{We propose a novel multivariate link function, $f$, for the parameters of the GEV density, defined as}
$$
(\psi_i, \tau_i,\phi_i,\gamma_i)\trp:=f(\mu_i,\sigma_i,\xi_i,\Delta_i)=(\log(\mu_i), \log(\sigma_i/\mu_i), h(\xi_i),d(\Delta_i))\trp\in\mathbb R^4,
$$ 
where 
$$
\gamma_i = d(\Delta_i) = {1\over2}\delta_0\{\log(\delta_0+\Delta_i)-\log(\delta_0-\Delta_i)\}
$$
with $\delta_0 = 0.008$. {The function $d$ and the selected value of $\delta_0$} ensure that $\Delta_i$ {lies} in the interval $(-0.008,0.008)$. {Most of the floods are observed in the period 1950 to 2013. The values of $\{1+\Delta_i(t-t_0)\}$ at the beginning and at the end of this period when $\Delta_i=0.008$ are $0.8$ and $1.3$ while $\Delta_i=-0.008$ gives the values $1.2$ and $0.7$. Thus, the interval for $\Delta_i$ allows a substantial change over this time period, and we believe that this interval is wide enough.} 
A Gaussian prior density with mean zero and standard deviation $0.5\delta_0$ is selected for $\gamma_i$.  
{Hence, there is a priori a $95$\% chance for $\gamma_i$ to be} in the interval  $(-0.008,0.008)$, which translates into a $95$\% prior chance {for $\Delta_i$ to be} in the interval  
$(-0.00609,0.00609)$. {This corresponds to about} $\pm 6$\% change per decade in the location parameter of the GEV distribution. {More details about the prior density for $\Delta$ are given in Section 1.4 in the Supplementary Material.}

\begin{figure}[t!]
\centering
\includegraphics[width=0.5\textwidth]{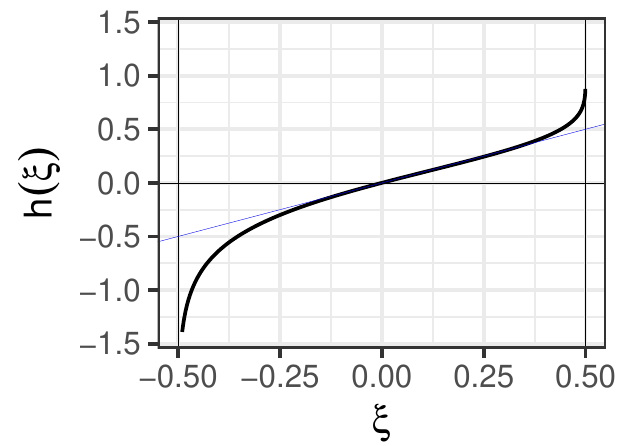}
\caption{The transformation for the shape parameter, $\phi=h(\xi)$ (black curve). The blue line is a reference line with intercept zero and slope one. The black vertical lines show the lower and upper bounds of $\xi$, $-0.5$ and $0.5$, respectively.}
\label{fig:transformation_phi}
\end{figure}

The transformed shape parameter is denoted by $\phi=h(\xi)$. {Our proposed choice for} the
function $h$ {relies on four natural criteria}: 
(1) the variance of the data distribution describing the floods (i.e., the GEV distribution) {should be} finite. This leads to the constraint $\xi < 0.5$;
(2) the upper bound of the GEV distribution {should not} be smaller than $\mu+2\sigma$.
This leads to the constraint $\xi > -0.5$;
(3) the asymptotic variance of the ML estimator of the transformed shape parameter $\phi$,
should vary as little as possible for values of $\phi$ that correspond to the interval $\xi \in [-0.3,0.3]$.
{This ensures that the estimation of parameters for the additive effects in the latent model for $\phi$ will be affected to a similar degree by the data from each site regardless of the underlying value of $\phi$, which is crucial when multiple covariates are involved in a linear way;}
(4) the transformation $h$ should be monotonic and such that $\phi$ is approximately equal to $\xi$ for values of $\xi$ around zero, which facilitates its interpretation. Mathematically, this means that {$\phi=h(\xi) \approx \xi$ when $\xi$ is close to zero}. Hence,
$$
{\textrm{d} h(\xi)\over\textrm{d}\xi}\bigg{|}_{\xi=0} =h'(0) = 1, \quad h(0)= 0.
$$
The constraints stemming from these four criteria motivate the transformation
$$
\phi = h(\xi) = a_\phi + b_\phi \log[-\log\{1-(\xi + 1/2)^{c_\phi}\}],
$$   
where $c_\phi=0.8$,
$$
b_\phi = -c_\phi^{-1}\log\{1 - (1/2)^{c_\phi}\}\{1 - (1/2)^{c_\phi}\}2^{c_\phi-1}=0.39563,   
$$
$$
a_\phi = -b_\phi \log[-\log\{1-(1/2)^{c_\phi}\}]=0.062376.
$$
Furthermore, the inverse of the transformation is 
\begin{equation}
\xi = h^{-1}(\phi) =g(\phi)= \bigg[1- \exp\bigg\{ -\exp\bigg({\phi - a_\phi\over b_\phi}\bigg)\bigg\}\bigg]^{1/c_\phi}-{1\over2}.
\label{eq:gtrans}
\end{equation}
The transformation from $\xi$ to $\phi$ via $h(\cdot)$ is shown in Figure \ref{fig:transformation_phi}. 
More details are given in {Sections~1.2 and 1.3 of} the Supplementary Material.

The generalized likelihood function for $\psi_i$, $\tau_i$, $\phi_i$ and $\gamma_i$ is the product of the likelihood function for site $i$ and the prior densities for $\phi_i$ and $\gamma_i$. Estimates based on the mode of our proposed generalized likelihood function will simply be referred to as ML estimates hereafter.   

\begin{figure}[t!]
\centering
\includegraphics[width=0.65\textwidth]{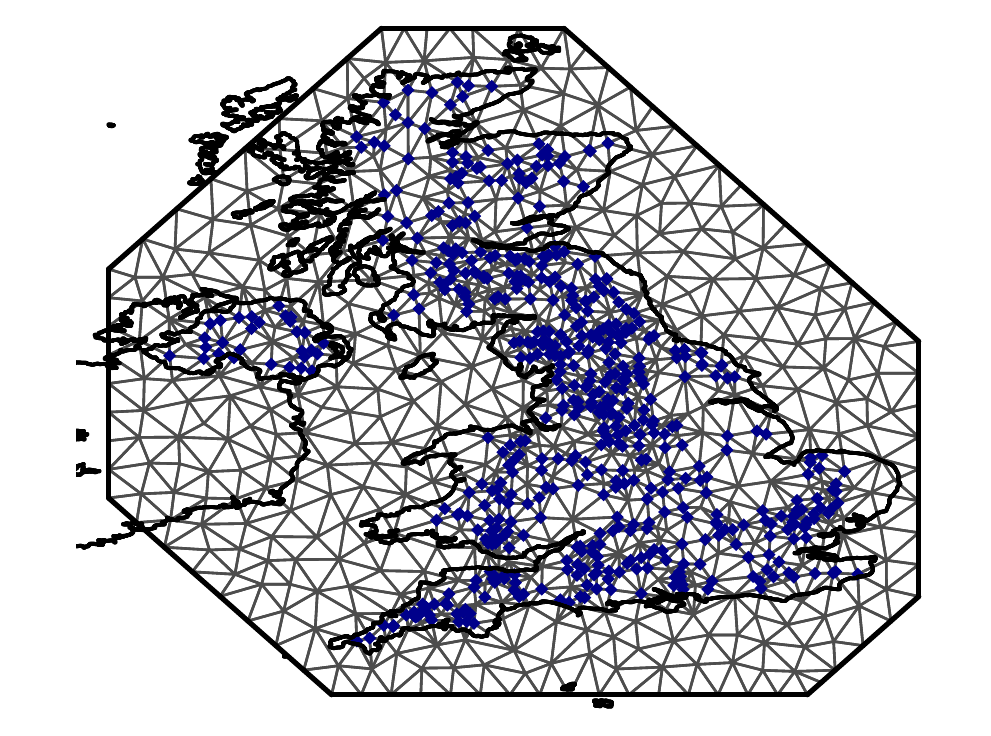}
\caption{Triangulated mesh over the UK used to define the SPDE spatial model components at the latent level. The blue dots denote the observational sites.}
\label{fig:mesh}
\end{figure}

At the latent level, our proposed regression models for the four transformed parameters $\psi$, $\tau$, $\phi$ and $\gamma$, 
may be expressed in {general} vector notation as
\begin{equation}
\label{eq:LatentLevel}
\begin{aligned}
\fat{\psi} &= X_{\psi} \fat{\beta}_{\psi}
+ A \fat{u}_{\psi} + \fat{\epsilon}_{\psi}, \\
\fat{\tau} &= X_{\tau} \fat{\beta}_{\tau}
+A \fat{u}_{\tau} + \fat{\epsilon}_{\tau}, \\
\fat{\phi} &= X_{\phi} \fat{\beta}_{\phi} {+A \fat{u}_{\phi}}+ \fat{\epsilon}_{\phi},\\
\fat{\gamma} &= X_{\gamma} \fat{\beta}_{\gamma} {+A \fat{u}_{\gamma}}+ \fat{\epsilon}_{\gamma},
\end{aligned}
\end{equation}
where $\fat{\psi} = (\psi_1,\ldots,\psi_J)\trp$, $\fat{\tau} = (\tau_1,\ldots,\tau_J)\trp$, $\fat{\phi} = (\phi_1,\ldots,\phi_J)\trp$ and $\fat{\gamma} = (\gamma_1,\ldots,\gamma_J)\trp$ are vectors of the model parameters.
The design matrices $X_{\psi}$, $X_{\tau}$, $X_{\phi}$ and $X_{\gamma}$ contain covariates, and $\fat{\beta}_{\psi}$, $\fat{\beta}_{\tau}$, $\fat{\beta}_{\phi}$ and $\fat{\beta}_{\gamma}$ are the corresponding coefficients. The vectors $\fat{u}_\psi$, {$\fat{u}_\tau$, $\fat{u}_\phi$ and $\fat{u}_\gamma$} are independent spatial random effects with Mat\'ern correlation structure, modeled using
finite-dimensional Gaussian Markov random fields (GMRF) via the stochastic partial differential equation (SPDE) approach of \cite{lindgren2011explicit}. 
A review of certain types of SPDE models can be found in \citet{bakka2018spatial} and detailed case studies in \cite{krainski2018advanced}.  
The SPDE approach requires a triangulated mesh defined over the region of interest, used to approximate the continuous-space random field. The mesh constructed for our application is illustrated in Figure \ref{fig:mesh}.
The matrix $A$ in \eqref{eq:LatentLevel} describes the projection from the nodes of the triangulation onto the observational sites.
The products, $A \fat{u}_\psi$, {$A \fat{u}_\tau$, $A \fat{u}_\phi$ and $A \fat{u}_\gamma$}, are thus the effects of the spatially structured model components at the observational sites. The spatial model components $\fat{u}_\psi$, {$\fat{u}_\tau$, $\fat{u}_\phi$ and $\fat{u}_\gamma$,} are governed by two hyperparameters each, namely, {the range parameters $\rho_l$, and the marginal standard deviations, $s_l$, for $l=\psi,\tau,\phi,\gamma$}. The smoothness parameter (usually denoted by $\nu$) of the Mat\'ern correlation structure is here fixed to one, in order to produce reasonably smooth realizations. 
The vectors $\fat{\epsilon}_{\psi}$, $\fat{\epsilon}_{\tau}$, $\fat{\epsilon}_{\phi}$ and $\fat{\epsilon}_{\gamma}$ in \eqref{eq:LatentLevel} contain unstructured zero-mean Gaussian model errors with variances $\sigma_{\epsilon\psi}^2,\sigma_{\epsilon\tau}^2$, $\sigma_{\epsilon\phi}^2$ and $\sigma_{\epsilon\gamma}^2$, respectively, sometimes called ``nugget effects'' in classical geostatistics{, which capture micro-scale variations}. That is,
$$
\fat{\epsilon}_{\psi}\sim\textrm{N}(0,\sigma_{\epsilon\psi}^2 I), \quad \fat{\epsilon}_{\tau}\sim\textrm{N}(0,\sigma_{\epsilon\tau}^2 I), \quad \fat{\epsilon}_{\phi}\sim\textrm{N}(0,\sigma_{\epsilon\phi}^2 I), \quad \fat{\epsilon}_{\gamma}\sim\textrm{N}(0,\sigma_{\epsilon\gamma}^2 I),
$$ 
where $I$ is the identity matrix. All these error terms are assumed to be mutually independent, both spatially and across parameters. 

Combining the four equations in \eqref{eq:LatentLevel} together, the model at the latent level may be rewritten more compactly as 
\begin{equation}
\label{eq:latLevel}
\fat{\eta}=Z \fat{\nu} + \fat{\epsilon},
\end{equation}
where
\begin{align*}
\fat{\eta} & = (\fat{\psi}\trp,\fat{\tau}\trp,\fat{\phi}\trp,\fat{\gamma}\trp)\trp,\\
\fat{\nu}  & = (\fat{\beta}_{\psi}\trp,\fat{u}_{\psi}\trp, \fat{\beta}_{\tau}\trp,\fat{u}_{\tau}\trp,\fat{\beta}_{\phi}\trp,{\fat{u}_{\phi}\trp,}\fat{\beta}_{\gamma}\trp,{\fat{u}_{\gamma}\trp})\trp,\\
\fat{\epsilon} & =(\fat{\epsilon}_\psi\trp,\fat{\epsilon}_{\tau}\trp,\fat{\epsilon}_{\phi}\trp,\fat{\epsilon}_{\gamma}\trp)\trp,\\
Z & =
\begin{pmatrix} 
X_\psi &  A_\psi & \fat{0} & \fat{0} & \fat{0} & \fat{0} & \fat{0} & \fat{0} \\
\fat{0} & \fat{0} & X_\tau & A_\tau & \fat{0} & \fat{0} & \fat{0} & \fat{0} \\
\fat{0} & \fat{0} & \fat{0} & \fat{0} & X_\phi & {A_\phi} & \fat{0} & \fat{0} \\
\fat{0} & \fat{0} & \fat{0} & \fat{0} & \fat{0} & \fat{0} & X_\gamma & {A_\gamma} \\
  \end{pmatrix},
\end{align*}
{where $\fat{0}$ denotes zero matrices of appropriate dimensions}.

{Prior specification for} the unknown parameters are such that the covariate coefficients, $\fat{\beta}_\psi$, $\fat{\beta}_\tau$, $\fat{\beta}_\phi$ and $\fat{\beta}_\gamma$, and the spatial model components, $\fat{u}_\psi$, {$\fat{u}_\tau$, $\fat{u}_\phi$ and $\fat{u}_\gamma$}, have  Gaussian priors while the hyperparameters, $\sigma_{\epsilon\psi}$, $\sigma_{\epsilon\tau}$, $\sigma_{\epsilon\phi}$, $\sigma_{\epsilon\gamma}$, $s_\psi$, {$s_\tau$,$s_\phi$, $s_\gamma$}, $\rho_\psi$, {$\rho_\tau$, $\rho_\phi$, $\rho_\gamma$}, are assigned penalized complexity (PC) priors to regularize the random effects of the model and shrink our complex Bayesian model towards a simpler reference model, thus preventing overfitting \citep{simpson2017penalising}. We use the same PC priors for $(s_\psi,\rho_\psi)^\top$, {$(s_\tau,\rho_\tau)^\top$, $(s_\phi,\rho_\phi)^\top$ and $(s_\gamma,\rho_\gamma)^\top$} as in \cite{fuglstad2017constructing}. 
Mutually independent prior densities are assumed for
$\sigma_{\epsilon\psi}$, $(s_\psi,\rho_\psi)^\top$, $\sigma_{\epsilon\tau}$, $(s_\tau,\rho_\tau)^\top$, $\sigma_{\epsilon\phi}$ and  $\sigma_{\epsilon\gamma}$.
The PC priors for $\sigma_{\epsilon\psi}$, $\sigma_{\epsilon\tau}$, $\sigma_{\epsilon\phi}$ and $\sigma_{\epsilon\gamma}$ are exponential distributions, while PC priors for $(s_\psi,\rho_\psi)^\top$, {$(s_\tau,\rho_\tau)^\top$, $(s_\phi,\rho_\phi)^\top$ and $(s_\gamma,\rho_\gamma)^\top$}, have joint densities
$$
\pi(s_l,\rho_l)=\lambda_{\rho,l} \lambda_{s,l} \rho_l^{-2}\exp(-\lambda_{\rho,l} \rho_l^{-1} -\lambda_{s,l} s_l), \ s_l>0, \ \rho_l>0, \  l \in \{\psi,\tau,{\phi,\gamma\}}, 
$$
{where $\lambda_{s,l}$ and $\lambda_{\rho,l}$ are calculated after selecting a threshold for $s_l$ and another threshold for $\rho_l$ that are believed to be a priori exceeded with probability $0.05$ and $0.95$, respectively; see \cite{fuglstad2017constructing}.}  

Prediction of flow at a set of ungauged sites involves predicting $\psi$, $\tau$, $\phi$ and $\gamma$, and requires covariates and coordinates of these sites.   
Modified versions of the equations in (\ref{eq:LatentLevel}) are needed to predict $\psi$, $\tau$, $\phi$ and $\gamma$. For example, let $\fat{\psi}_{\textrm{un}}$ 
denote $\psi$ at the ungauged sites. Its formula is 
$$
\fat{\psi}_{\textrm{un}} = X_{\psi,\textrm{un}} \fat{\beta}_\psi + A_{\textrm{un}} \fat{u}_\psi + \fat{\epsilon}_{\psi,\textrm{un}},
$$
where $X_{\psi,\textrm{un}}$ contains the covariates, $A_{\textrm{un}}$ is such that $A_{\textrm{un}} \fat{u}_\psi$
is a prediction of the spatial component at the ungauged sites, and $\fat{\epsilon}_{\psi,\textrm{un}}$ contains the unstructured random effects of the ungauged sites which all have 
mean zero and variance $\sigma^2_{\epsilon\psi}$. 

{Further details about the statistical model can be found in Section 1.5 in the Supplementary Material.} In the next section, we describe---using the same notation---how to fit our proposed extended LGM to potentially very large datasets, by exploiting and extending an approximate Bayesian inference scheme {called Max-and-Smooth}.

\section{Approximate posterior inference}
\label{ch:inference}
\subsection{Gaussian-based approximation to the posterior density}
\label{ch:GaussianApprox}

To deal with the high dimensionality of the data and the latent parameters, and to enable model selection at the latent level, a fast computational method for posterior inference
is required. \citet{hrafnkelsson2020approx} recently developed
a general approximate posterior inference scheme for extended LGMs, called Max-and-Smooth. {While \citet{hrafnkelsson2020approx} tested the performance of Max-and-Smooth in a simple Gaussian regression example with spatially-varying mean and log-variance defined on a lattice, we here extend this methodology and validate it in our extreme-value framework with a heavy-tailed non-Gaussian likelihood. It is the first time that this methodology is rigorously tested in such a large-scale, strongly non-stationary and highly non-Gaussian data setting.}  
This approach is based on a Gaussian approximation to the likelihood function,
which leads to a Gaussian--Gaussian pseudo model that is inferred instead of the initial extended LGM. This approximation is very accurate when numerous time replicates are available {and/or when the likelihood is close to Gaussian, and it} significantly facilitates inference, while also substantially reducing the computational burden. 

To be more precise, let $\fat{\theta}$ be the vector containing the hyperparameters of the extended latent Gaussian model in Section \ref{ch:mdlSpec}, namely,  
$$
\fat{\theta}=(\theta_1,\ldots,\theta_{12})\trp = (\sigma_{\epsilon\psi}, s_{\psi}, \rho_{\psi},  \sigma_{\epsilon\tau}, s_{\tau},\rho_{\tau}, \sigma_{\epsilon\phi}, {s_{\phi},\rho_{\phi}, \sigma_{\epsilon\gamma}, s_{\gamma},\rho_{\gamma}})\trp.
$$
Then, the posterior density can be written as
\begin{equation}
\begin{aligned}
\pi(\fat{\eta},\fat{\nu},\fat{\theta}\mid\fat{y}) 
& \propto  \pi(\fat{\theta})
\pi(\fat{\nu}\mid\fat{\theta})
\pi(\fat{\eta}\mid\fat{\nu},\fat{\theta})
\pi(\fat{y}\mid\fat{\eta}) \\
& \propto  \pi(\fat{\theta})
\pi(\fat{\nu}\mid\fat{\theta})
\pi(\fat{\eta}\mid\fat{\nu},\fat{\theta})
L(\fat{\eta}\mid\fat{y}),
\end{aligned}\label{eq:1}
\end{equation}
where $L(\cdot\mid\cdot)$ denotes the generalized likelihood function. We then approximate the generalized likelihood function by a Gaussian density with mean $\hat{\fat{\eta}}$, and covariance matrix $\Sigma_{\eta y}=Q_{\eta y}^{-1}$, where $\hat{\fat{\eta}}$ is the mode of the generalized log-likelihood function, $\log L$, and $Q_{\eta y}^{-1}$ is the inverse of the negative Hessian matrix of $\log L$ evaluated at $\hat{\fat{\eta}}$. Let $\hat{L}(\fat{\eta}\mid\fat{y})$ be the Gaussian approximation to the generalized likelihood function. Denoting the approximation to the posterior density by $\hat{\pi}(\cdot\mid\cdot)$, we have
\begin{equation}
\begin{aligned}
\hat{\pi}(\fat{\eta},\fat{\nu},\fat{\theta}\mid\fat{y}) 
& \propto  \pi(\fat{\theta})
\pi(\fat{\nu}\mid\fat{\theta})
\pi(\fat{\eta}\mid\fat{\nu},\fat{\theta})
\hat{L}(\fat{\eta}\mid\fat{y}) \\
& \propto   \pi(\fat{\theta})
\pi(\fat{\nu}\mid\fat{\theta})
\pi(\fat{\eta}\mid\fat{\nu},\fat{\theta})
\textrm{N}(\fat{\eta}\mid\hat{\fat{\eta}},Q_{\eta y}^{-1}), 
\end{aligned}\label{eq:2}
\end{equation}
where $\textrm{N}(\cdot\mid\cdot)$ denotes a Gaussian density.
Now, consider a pseudo model that is such that $\hat{\fat{\eta}}$ (obtained by maximizing $L(\cdot\mid\fat{y})$ at each site separately in a first step) is treated as the data (i.e., as noisy measurements of $\fat{\eta}$), and assume that $\fat{\eta}$ is modeled with the linear model in \eqref{eq:latLevel} used at the latent level of our original model. The proposed data density is now $\pi(\fat{\hat{\eta}}\mid\fat{\eta})=\textrm{N}(\hat{\fat{\eta}}\mid\fat{\eta},Q_{\eta y}^{-1})$ where $Q_{\eta y}$ is as above and assumed to be known. The posterior density for this model can thus be written as 
\begin{equation}
\begin{aligned}
\pi(\fat{\eta},\fat{\nu},\fat{\theta}\mid\fat{\hat{\eta}}) & 
\propto \pi(\fat{\theta})\pi(\fat{\eta},\fat{\nu}\mid\fat{\theta})
\pi(\fat{\hat{\eta}}\mid\fat{\eta})\\
&\propto \pi(\fat{\theta})\pi(\fat{\eta},\fat{\nu}\mid\fat{\theta})\textrm{N}(\fat{\hat{\eta}}\mid\fat{\eta},Q^{-1}_{\eta y}) \\
&\propto \pi(\fat{\theta})\pi(\fat{\eta},\fat{\nu}\mid\fat{\theta})\hat{L}(\fat{\eta}\mid\fat{\hat{\eta}}), 
\end{aligned}\label{eq:3}
\end{equation}
where $\hat{L}(\fat{\eta}\mid\fat{\hat{\eta}})=\hat{L}(\fat{\eta}\mid\fat{y})$.
Hence, the above posterior density in \eqref{eq:3} is the same as the Gaussian approximation \eqref{eq:2} to the \emph{exact} posterior density \eqref{eq:1}. This observation motivates a two-step approximate Bayesian inference scheme, called Max-and-Smooth in \citet{hrafnkelsson2020approx}, whereby the estimates $\fat{\hat\eta}$ are obtained first at each site separately (Max) and are then smoothed by inferring the Gaussian--Gaussian pseudo model (Smooth). The main reason for using the pseudo model is that it is fully conjugate, 
and thus posterior inference takes less computation time than posterior inference for the exact model.
Notice that once the estimates $\hat{\fat{\eta}}$ are obtained, the computational time required for fitting the pseudo model no longer depends on the number of time replicates. Therefore, Max-and-Smooth benefits from large datasets with a lot of time replicates in two ways: the Gaussian approximation to the likelihood function not only becomes more accurate with more time replicates, but the relative computational cost with respect to an ``exact'' MCMC approach also decreases dramatically. We also emphasize here that although our inference approach is done in two consecutive steps (i.e., first estimating $\hat{\fat{\eta}}$, and second inferring $\fat{\eta}$ and $\fat{\theta}$ with the Gaussian--Gaussian pseudo model for $\hat{\fat{\eta}}$), the uncertainty involved in the first step is properly propagated into the second step in a way that provides a valid approximation to the full posterior distribution. {Further details about the approximate posterior inference can be found in Section 1.6 in the Supplementary Material.}

{We assess} the accuracy of the Gaussian approximation to $i$-th likelihood contribution $L(\fat{\eta}_i\mid\fat{y}_i)$, where $L(\fat{\eta}\mid\fat{y})=\prod_{i=1}^{J}L(\fat{\eta}_i\mid\fat{y}_i)$, in {Section 1.7 of} the Supplementary Material for various parameter settings when using the GEV distribution without the time trend. {Expectedly, our} results show that the approximation becomes more accurate as the number of time replicates $T$ increases. It is exceedingly accurate for the transformed location and shape parameters $\psi_i$ and $\phi_i$ when $T=50$ or $T=80$ but has a slight negative bias for the transformed scale parameter $\tau_i$ in these temporal dimensions or for smaller values of $T$.

\subsection{Model selection procedure based on INLA}
\label{ch:mdlSel}

{In order to quickly select covariates and spatial model components to include in the linear regression models for the transformed parameters at the latent level, an additional approximation may be performed. This approximation, detailed below, is only used to guide model selection at a preliminary stage in a data-driven way that respects the core model structure, but it should not be used in the main inference procedure to fit the final model that has been selected. More precisely, for} model selection purposes only, the Gaussian--Gaussian pseudo model {may be} further simplified by setting the off-diagonal elements
of $Q^{-1}_{\eta y}$ equal to zero, i.e., {by neglecting all covariances between the parameters $(\hat{\psi}_i, \hat{\tau}_i, \hat{\phi}_i, \hat{\gamma}_i)\trp$ for each catchment $i$, and only considering the marginal variances.} The simplified version of the data density is thus 
$$
\pi(\fat{\hat{\eta}}\mid\fat{\eta}, \fat{\nu}, \fat{\theta}) = \textrm{N}(\fat{\hat{\eta}}\mid\fat{\eta},\text{diag}(Q_{\eta y}^{-1})),
$$
while the pseudo model for $\fat{\hat{\psi}}$ at the data level then takes the form 
$$
\pi(\fat{\hat{\psi}}\mid\fat{\psi}, \fat{\nu}, \fat{\theta}) = \textrm{N}(\fat{\hat{\psi}}\mid\fat{\psi},Q_{\psi y}^{-1}),
$$
where $Q_{\psi y}^{-1}$ is the sample variance of $\hat{\fat{\psi}}$ obtained from $\text{diag}(Q_{\eta y}^{-1})$. Similarly, the pseudo models for $\fat{\hat{\tau}}$, $\fat{\hat{\phi}}$ and $\fat{\hat{\gamma}}$ at the data level take the forms 
\begin{equation}
\begin{aligned}
\pi(\fat{\hat{\tau}}\mid\fat{\tau}, \fat{\nu}, \fat{\theta}) &= \textrm{N}(\fat{\hat{\tau}}\mid\fat{\tau},Q_{\tau y}^{-1}), \\ \pi(\fat{\hat{\phi}}\mid\fat{\phi}, \fat{\nu}, \fat{\theta}) &= \textrm{N}(\fat{\hat{\phi}}\mid\fat{\phi},Q_{\phi y}^{-1}), \\
\pi(\fat{\hat{\gamma}}\mid\fat{\gamma}, \fat{\nu}, \fat{\theta}) &= \textrm{N}(\fat{\hat{\gamma}}\mid\fat{\gamma},Q_{\gamma y}^{-1}) .
\end{aligned}
\nonumber
\end{equation}
This translates into assuming independence between the transformed location intercept, scale, shape and time trend parameters at the ``data level'' of the pseudo model 
such that the linear model for each of these four parameters can be
inferred separately. This separation allows applying the very fast integrated nested Laplace approximation (INLA) methodology \citep{rue2009approximate} and exploiting the associated \texttt{R} package to estimate models independently for each of the four parameters, thus making a $10$-fold
cross-validation feasible. The INLA methodology bypasses MCMC sampling and the associated convergence assessment issues for LGMs with univariate link functions, by relying on a fast numerical approximation to the posterior density, which is similar in spirit (albeit with some important differences) to our proposed procedure detailed in Section~\ref{ch:GaussianApprox} in the sense that INLA also exploits Gaussian-based approximations; for more details on INLA and its application to spatial models, see \citet{bakka2018spatial}, and for an example of INLA applied to extreme precipitation data, see \citet{Opitz.etal:2018}. 

{In Section~6 in the Supplementary Material, we show that inferring $\psi$, $\tau$ and $\phi$ separately for the purpose of model selection based on three separate models can indeed be justified, since the posterior means and the marginal posterior intervals of the $\beta$s and the $u$s, for say $\psi$, are numerically similar when separate INLA runs are used (assuming incorrectly that the parameters are independent in the posterior) or when a correct full MCMC approach is used. In other words, information about the $\beta$s and $u$s for $\psi$ is found in $\hat\psi$, while the other parameters (e.g., $\hat\tau$) add negligible knowledge about $\psi$ when $\hat\psi$ is already known. However, these independent INLA fits neglect the dependence between parameters, and thus cannot be used to evaluate posterior uncertainty in return levels, which involve the GEV quantile function that combines all parameters, $\psi$, $\tau$, $\phi$ and $\gamma$. Neglecting the posterior covariance between $\hat\psi$, $\hat\tau$, $\hat\phi$ and $\hat\gamma$ can thus dramatically impact the posterior variability of return levels. Moreover, it turns out that $\gamma$ is quite sensitive to ignoring the dependence in the likelihood function. In particular, the posterior means of $\gamma$ have a wider spread with the INLA approach than with the full correct model, and the corresponding posterior standard deviations are larger when computed with INLA. Thus, the model for $\gamma$ that is selected by the INLA approach needs be interpreted with caution. For further details, see Section 6 in the Supplementary Material.}  

The separate inferences for $\psi$, $\tau$, $\phi$ and $\gamma$ can be made very efficiently using INLA, taking 2, 2, 1, and 1, minutes, respectively, on a standard laptop {for our large dataset}. It is therefore possible to select the ``best'' covariates by cross-validation using a forward selection procedure in a reasonable amount of time.
The $10$-fold cross-validation procedure that we used involves splitting the data into ten disjoint randomly chosen subsets with the same number of sites, the data from each site being used only once. The model is then trained on nine of those subsets and then used to predict the left-out subset and to compute the corresponding root mean squared prediction error, defined as the sum of squared differences between ML estimates of say $\fat{\psi}$ at the left-out sites and predictions of these based on the $X_\psi \fat{\beta}_\psi+A_\psi \fat{u}_\psi$ part of the model using the posterior mean of $\fat{\beta}_\psi$ and $\fat{u}_\psi$. This step is repeated until each of the ten subsets have been used for testing, and the root mean squared prediction errors of each test set are then averaged to calculate the overall root mean squared prediction error (also simply called the ``test error'') for a given model. In our application, the whole model selection procedure took only about 7 hours in total for all four parameters. As a result, linear models with good prediction properties were found for each of the four parameters. Moreover, in order to further motivate the inclusion of the spatial model components, the forward model selection was performed with and without the spatial model components for each of $\fat{\psi}$, $\fat{\tau}$, $\fat{\phi}$ and $\fat{\gamma}$. 

\section{Results}

\label{ch:results}
\subsection{Model selection results}
\label{ch:results_mod_sel}

\begin{figure}[t!]
\centering
\includegraphics[width=0.9\textwidth]{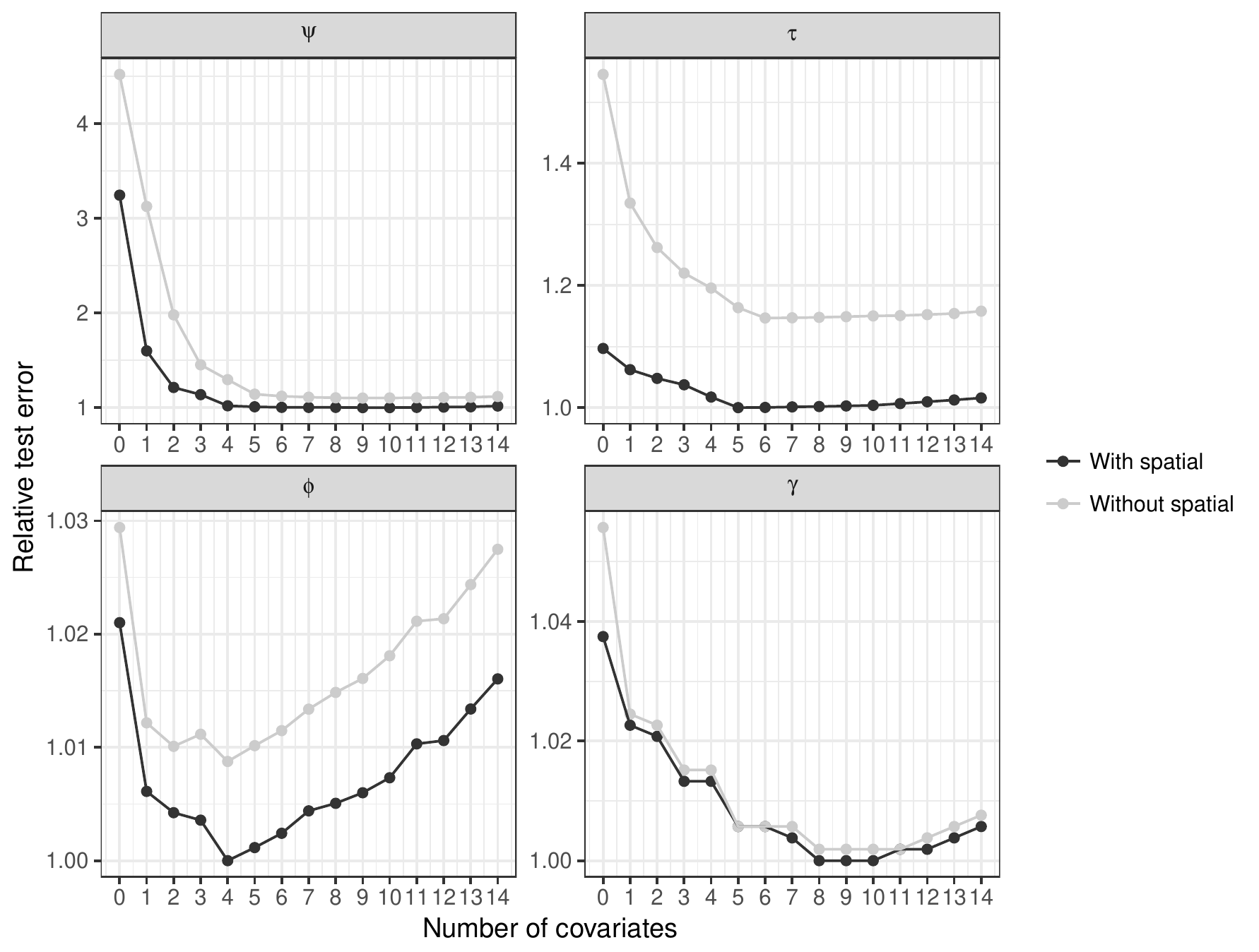}
\caption{Relative test error (i.e., the overall root mean squared prediction error rescaled by that of the best fitted model) plotted as a function of the number of covariates selected by the forward selection procedure in the models for $\fat{\psi}$, $\fat{\tau}$, $\fat{\phi}$ and $\fat{\gamma}$, upper left to lower right, respectively. Black (respectively grey) curves correspond to models including (respectively not including) spatially structured model components at the latent level.} \label{fig:spatialForward}
\end{figure}

We applied the forward model selection approach based on INLA described in Section \ref{ch:mdlSel} to select covariates for the linear models of $\fat{\psi}$, $\fat{\tau}$, $\fat{\phi}$ and $\fat{\gamma}$ that are given by Eq. (\ref{eq:LatentLevel}). Figure \ref{fig:spatialForward} shows the relative test error (i.e., the overall root mean squared prediction error rescaled by that of the best fitted model) for the pseudo models fitted separately to $\fat{\hat\psi}$, $\fat{\hat\tau}$, $\fat{\hat\phi}$ and $\fat{\hat\gamma}$ as a function of the number of covariates selected by the forward selection procedure. 

When selecting a final model for each of the four parameters based on the overall root mean square prediction error, we opt for a parsimonious model with good prediction properties. Due to the high complexity of the spatial component we only add it to a final model if the gain is considerable.
In particular, the overall root mean square prediction error is significantly lower when spatial components are included in $\fat{\psi}$ and $\fat{\tau}$, which suggests that they should be included in the final model, while this is not the case for $\fat{\phi}$ and $\fat{\gamma}$. 
Based on Figure \ref{fig:spatialForward} we selected models for $\fat{\psi}$, $\fat{\tau}$, $\fat{\phi}$ and $\fat{\gamma}$ with four, five, one and one covariate, respectively.
There is a negligible improvement in the overall root mean square prediction error for models with additional covariates.   
All models also include an intercept. Table \ref{tab:covariatesSelected} shows the model components (covariates and spatial effects) selected in the final models for $\fat{\psi}$, $\fat{\tau}$, $\fat{\phi}$ and $\fat{\gamma}$. 
\begin{table}[t!]
\rowcolors{2}{gray!6}{white}
\caption{\label{tab:covariatesSelected}
The latent level of our final model.
Model components (covariates or spatial effects) selected in our final model are indicated by check marks (\checkmark) for $\boldsymbol{\psi}$, $\boldsymbol{\tau}$,  $\boldsymbol{\phi}$ and $\boldsymbol{\gamma}$.}
    \centering
    \begin{tabular}{lcccc}
    \toprule
\textbf{Model component}  & $\boldsymbol{\psi}$ \textbf{model} & $\boldsymbol{\tau}$ \textbf{model} & $\boldsymbol{\phi}$ \textbf{model} & $\boldsymbol{\gamma}$ \textbf{model}  \\
\midrule
\showrowcolors
       $\log(\text{AREA})$ & \checkmark & \checkmark & & \\
       $\log(\text{SAAR})$ & \checkmark & \checkmark & & \\
       $\log(\text{FARL})$ & \checkmark & \checkmark & & \\
       $\log(\text{FPEXT})$ &  & \checkmark & \checkmark & \\
       $\log(\text{URBEXT} + 1)$ &  & \checkmark & & \\
       $\text{BFIHOST}^2$ & \checkmark &  & & \\
       $\log(\text{PROPWET})$ &  &  & & \checkmark \\
       Spatial component & \checkmark & \checkmark & & \\
\bottomrule
    \end{tabular}
    \rowcolors{2}{white}{white}
\end{table}

\subsection{Parameter estimates and interpretation}

\begin{figure}[t!]
\centering 
\includegraphics[width=0.495\linewidth]{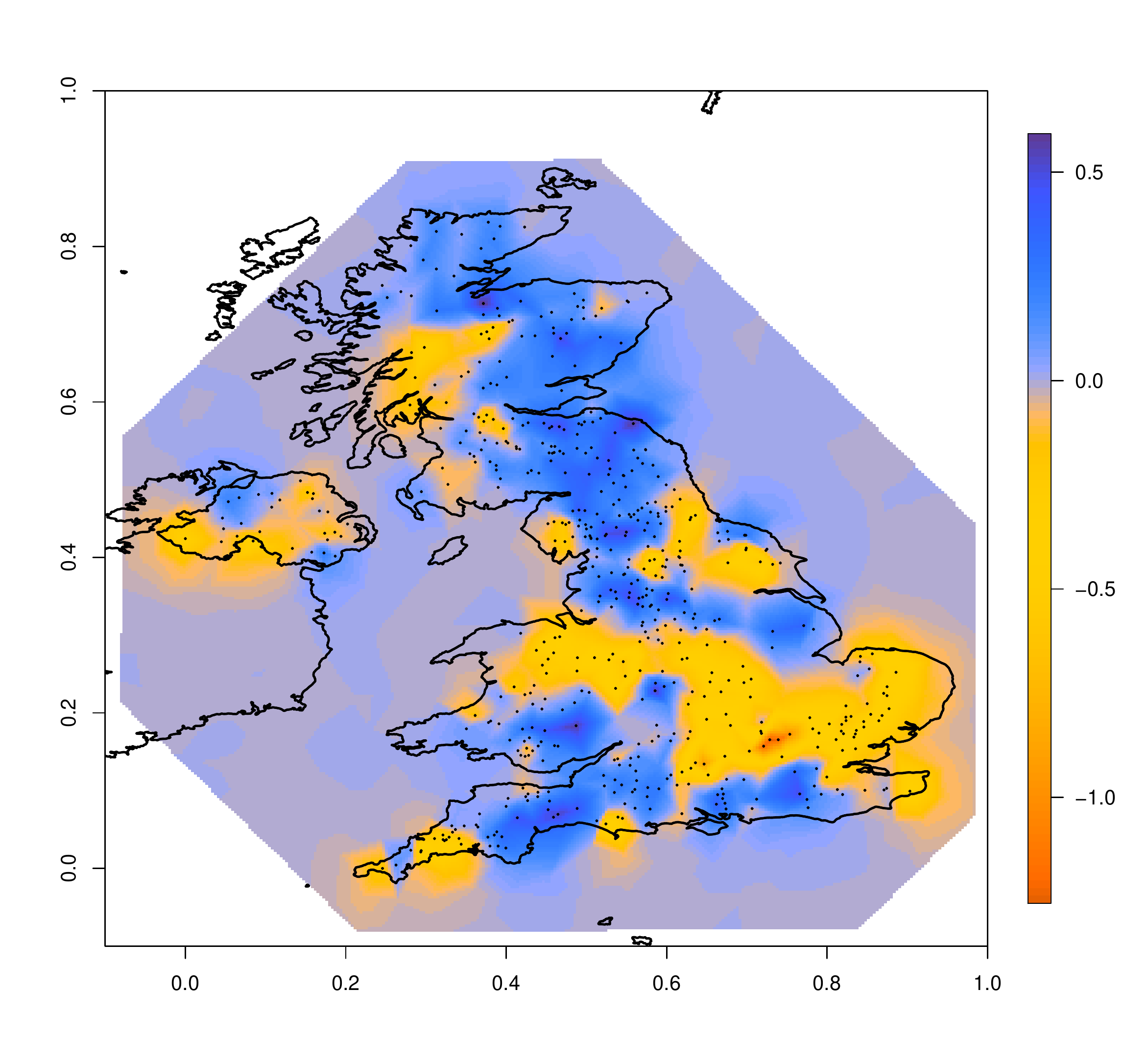}
\includegraphics[width=0.495\linewidth]{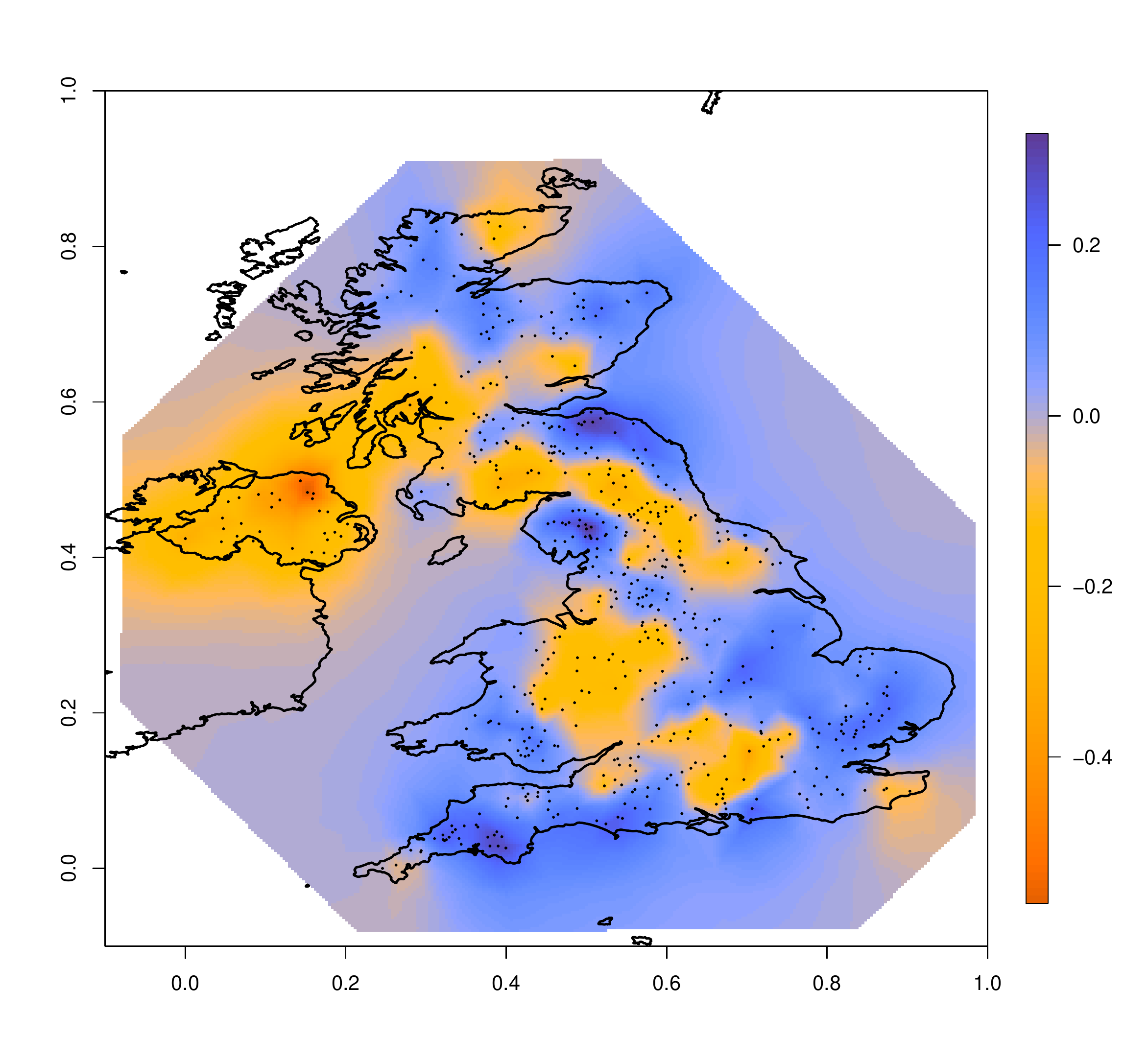}
\caption{The posterior mean of the spatial model components of $\fat{\psi}$ (left) and $\fat{\tau}$ (right). Note that the color scales are different in both panels, although zero has the same color.}
\label{fig:spatialpsitau}
\end{figure}

Summary statistics for the estimated hyperparameters are presented in {Section 3.2 of} the Supplementary Material. {Convergence assessment of the posterior simulation for the full model is given in Section 3.1 of the Supplementary Material.} 
A comparison of the marginal standard deviations of the spatial model components and the unstructured model components of $\fat{\psi}$ and $\fat{\tau}$ shows that the marginal standard deviations of the spatial model components are bigger than those of the random errors $\fat{\epsilon}_\psi$ and $\fat{\epsilon}_\tau$, which suggests that the inclusion of latent spatial random effects helps to borrow strength across locations. Fitting simplified models to $\fat{\psi}$ and $\fat{\tau}$ that do not contain spatial model components but use the same covariates as in the full model, resulted in the estimates $\sigma_{\psi\epsilon}^* = 0.397$ and $\sigma_{\tau\epsilon}^* = 0.211$. Here, $\sigma_{\psi\epsilon}^*$ and $\sigma_{\tau\epsilon}^*$ denote the posterior means of the standard deviations of the unstructured model components in the simplified models for $\fat{\psi}$ and $\fat{\tau}$ without the spatial component. This result is not surprising since 
$$
0.397=\sigma_{\psi\epsilon}^* \approx \sqrt{s_{\psi,\text{post}}^2+\sigma_{\psi \epsilon,\text{post}}^2} = \sqrt{0.311^2+0.247^2}=0.397
$$
and
$$
0.211 = \sigma_{\tau\epsilon}^* \approx \sqrt{s_{\tau, \text{post}}^2+\sigma_{\tau \epsilon, \text{post}}^2} = \sqrt{0.182^2+0.133^2}=0.225,
$$
where $s_{\psi,\text{post}}^2$, $\sigma_{\epsilon\psi,\text{post}}^2$, $s_{\tau,\text{post}}^2$ and $\sigma_{\epsilon\tau,\text{post}}^2$ denote the posterior means of $s_{\psi}^2$, $\sigma_{\epsilon\psi}^2$, $s_{\tau}^2$ and $\sigma_{\epsilon\tau}^2$, respectively,
in the final model with the spatial model components. 
Thus, the calculations above show that the sum of the marginal variances of the spatial model components and the variances of the unstructured model components in the spatial model was approximately equal to the variances of the unstructured error terms in the model that did not have spatial model components. This suggests that the spatial model components explain variability that would otherwise not have been explained by the covariates.   
Figure \ref{fig:spatialpsitau} displays the posterior mean of the spatial model components of $\fat{\psi}$ and $\fat{\tau}$. 
The spatial model component of $\fat{\psi}$ takes values mainly between $-0.5$ and $0.5$, which corresponds to a multiplicative factor for $\mu$ that varies spatially between $0.61$ and $1.65$. Similarly, the spatial model component of $\fat{\tau}$ takes values mainly between $-0.3$ and $0.3$ which translates into a multiplicative factor for $\sigma$ that varies spatially between $0.74$ and $1.35$ after taking $\mu$ into account. The spatial model component of $\fat{\tau}$ has a range parameter that is greater than the one of $\fat{\psi}$ by a factor two. This indicates that there is a stronger spatial smoothing for the transformed scale parameter $\fat{\tau}$, and the effect of this can indeed be seen in Figure \ref{fig:spatialpsitau}. The spatial components represent the effect of location not explained by the covariates. When $u_\psi$ is positive, then the joint effect of the covariates is not large enough in the sense that extreme flow at these sites is on average higher than the covariates would have predicted without the spatial component. 
The average annual rainfall (SAAR) has the strongest spatial structure of the covariates, thus, to some extent the spatial components of $\fat{\psi}$ and $\fat{\tau}$ are compensating for SAAR as a predictor for the GEV location and scale parameters of the extreme flow.

The original GEV location intercept and scale parameters, $\fat{\mu}$ and $\fat{\sigma}$, can be written in terms of $\fat{\psi}$ and $\fat{\tau}$ for station $i$ as 
$$
\mu_i = \exp(\psi_i) = \exp(\fat{x}_{\psi,i}\fat{\beta}_\psi + \fat{a}_i \fat{u}_\psi + \epsilon_{\psi,i})
$$
and
$$
\sigma_i = \exp(\psi_i+\tau_i) = \exp(\fat{x}_{\psi,i}\fat{\beta}_\psi + \fat{x}_{\tau,i}\fat{\beta}_\tau + \fat{a}_i(\fat{u}_\tau + \fat{u}_\psi) + \epsilon_{\psi,i} + \epsilon_{\tau,i}).
$$
Using the posterior means as point estimates for the covariates coefficients, the estimated location intercept, scale, shape and time trend parameters may be expressed as
\begin{equation*}
\label{eq:psiCovariates}
\begin{aligned}
\mu_i =& e^{-11.622}(\textrm{AREA}_i)^{0.892}(\textrm{SAAR}_i)^{1.661}(\textrm{FARL}_i)^{3.616} \\
& \times e^{-3.272(\textrm{BFIHOST}_i)^2} 
 e^{\fat{a}_i\fat{u}_\psi + \epsilon_{\psi,i}},
\end{aligned}
\end{equation*}
\begin{equation*}
\label{eq:sigmCovariates}
\begin{aligned}
\sigma_i =& e^{-8.115}(\textrm{AREA}_i)^{0.837}(\textrm{SAAR}_i)^{0.981}(\textrm{FARL}_i)^{2.578}
 \\
& \times (\textrm{URBEXT}_i+1)^{-0.750}  (\textrm{FPEXT}_i)^{-0.135} e^{\fat{a}_i(\fat{u}_\psi+\fat{u}_\tau) + \epsilon_{\psi,i}+\epsilon_{\tau,i}},
\end{aligned}
\end{equation*}
\begin{equation*}
\label{eq:xiCovariates}
\xi_i = \bigg[1- \exp\bigg\{ -\exp\bigg({-0.102-0.037\log(\textrm{FPEXT}_i) + \epsilon_{\phi,i} - a_\phi\over b_\phi}\bigg)\bigg\}\bigg]^{1/c_\phi}-{1\over 2},
\end{equation*}
\begin{equation*}
\label{eq:deltaCovariates}
\Delta_i = 2\delta_0 
\bigg[1+\exp\bigg\{{-2\delta_0^{-1}(\beta_{\gamma,1}+\beta_{\gamma,2}\log(\textrm{PROPWET}_i)+\epsilon_{\gamma,i})}\bigg\}\bigg]^{-1} - \delta_0,
\end{equation*}
where $a_\phi=0.062376$, $b_\phi=0.39563$, $c_\phi=0.8$, $\delta_0=0.008$, $\beta_{\gamma,1}=3.09\cdot 10^{-3}$ and $\beta_{\gamma,2}=1.97\cdot 10^{-3}$.
The effect of a covariate that is used in both the $\fat{\psi}$ model and the $\fat{\tau}$ model on $\fat{\sigma}$ needs to be handled in a particular way; for example, the effect of AREA on $\sigma_i$ is 
governed by the exponent $\beta_{\psi,2}+\beta_{\tau,2}$
which is estimated to be $0.837$ since the posterior estimates of $\beta_{\psi,2}$ and $\beta_{\tau,2}$ are $0.892$ and $-0.055$, respectively.

\begin{figure}[t!]
\centering
\includegraphics[width=0.8\textwidth]{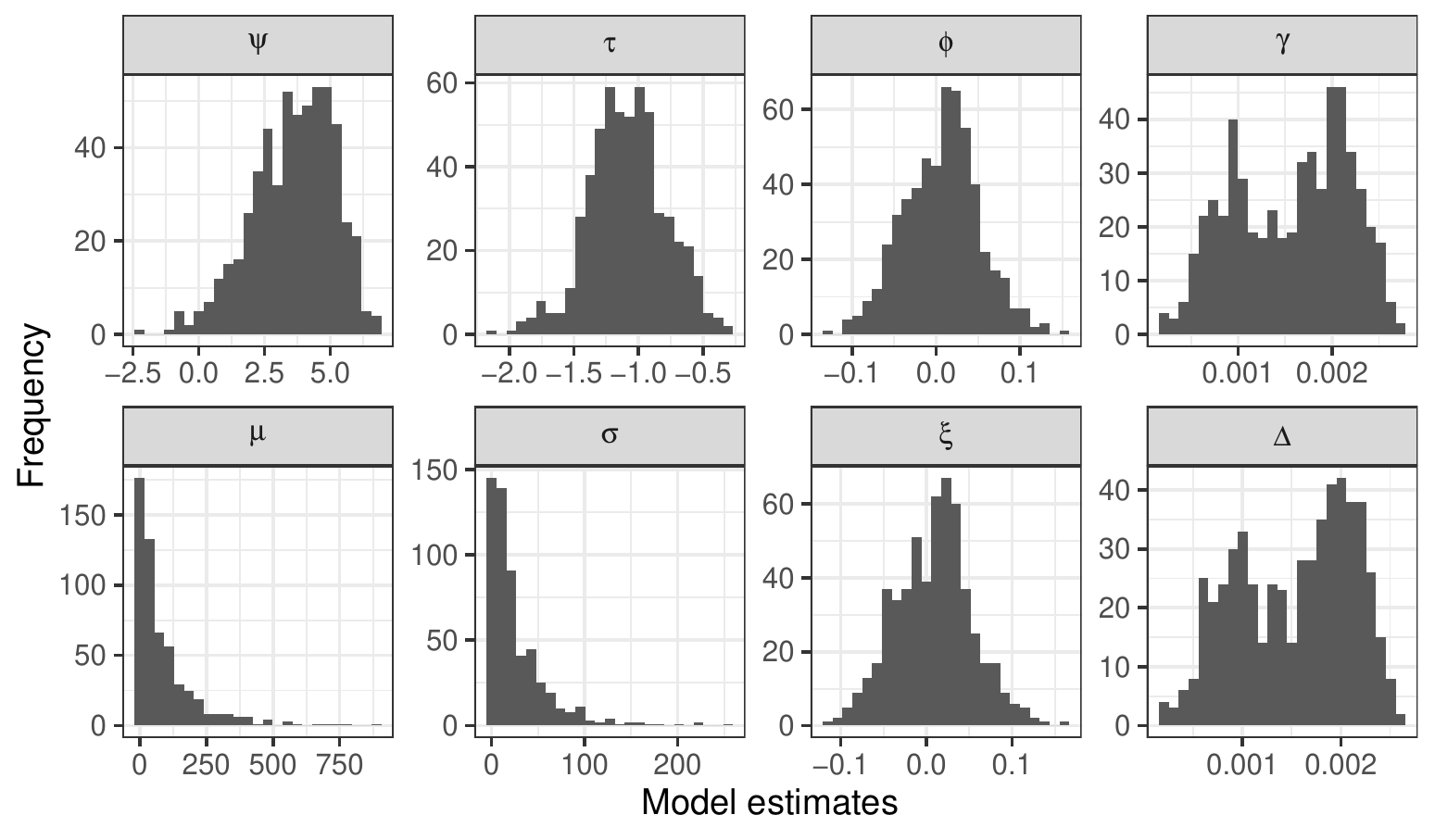}
\caption{Histograms of the posterior means of the GEV parameters, $\mu$, $\sigma$ and $\xi$, and the time trend $\Delta$ (bottom), from each of the observational sites along with the posterior means of the transformed parameters, $\psi$, $\tau$, $\phi$ and $\gamma$ (top).} \label{fig:modelEstimatesHistograms}
\end{figure}

Figure \ref{fig:modelEstimatesHistograms} shows histograms of the posterior means of the GEV parameters, $\mu_i$, $\sigma_i$, $\xi_i$ and $\Delta_i$, from all observational sites $i$ along with the posterior means of the corresponding transformed parameters, $\psi_i$, $\tau_i$, $\phi_i$ and $\gamma_i$. The histograms of the preliminary site-wise ML estimates of the same parameters from these sites are shown in {Section 2.1 of} the Supplementary Material.
Based on the posterior means of the shape parameters, the data appear to be close to the Gumbel distribution with light tails as $\xi_i\approx0$ at all sites $i$. As for the time trend $\Delta_i$, most sites indicate a {slight} positive increase of about $1.5\%$ per decade, revealing a temporal effect on river flow extremes in the United Kingdom that may be due to climate change. 
\begin{figure}[t!]
\centering
\includegraphics[width=0.8\textwidth]{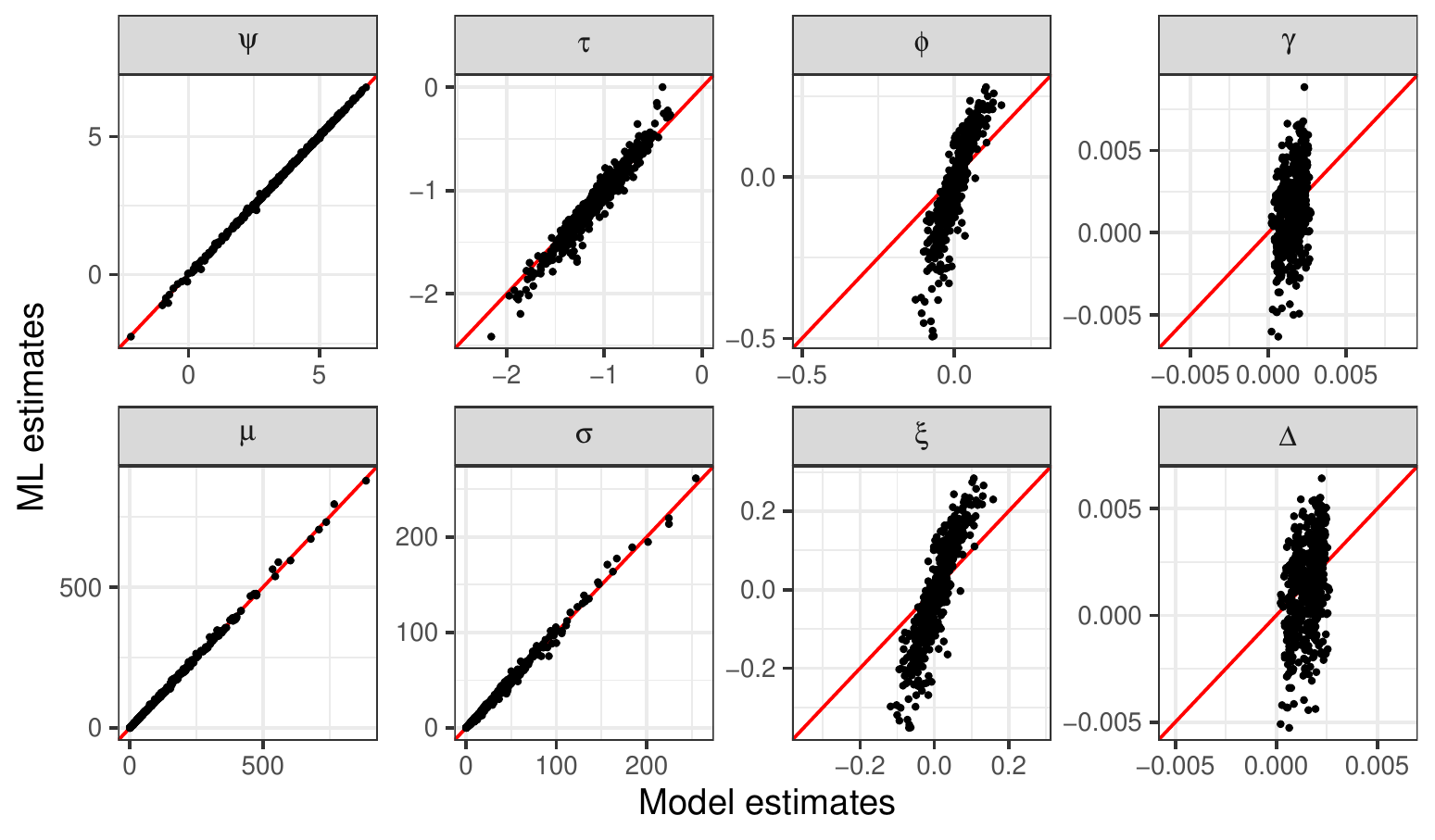}
\caption{The ML estimates of the GEV parameters, $\mu$, $\sigma$ and $\xi$, and the time trend parameter $\Delta$ (bottom), and their transformations, $\psi$, $\tau$, $\phi$ and $\gamma$ (top), as a function of their posterior estimates.} \label{fig:MLEstimatesVSModelEstimates}
\end{figure}
Figure \ref{fig:MLEstimatesVSModelEstimates} shows the preliminary site-wise ML estimates of the GEV parameters and the transformed GEV parameters plotted against their final posterior means. The values of $\mu_i$ and $\sigma_i$ vary a lot across catchments. Despite these large differences, our flexible spatial model can accurately capture the data distribution over space thanks to our choice of informative covariates and the inclusion of latent spatial model components and unstructured model components. There is an excellent match between the preliminary site-wise ML estimates of the location parameter $\mu_i$ and their final model-based counterparts, and similarly (albeit to a less degree) for the scale parameter $\sigma_i$, and yet the uncertainty of the posterior estimates is substantially lower than the uncertainty associated with the ML estimates. This is due to spatial smoothing, and it implies that our Bayesian spatial model adequately borrows strength across locations for improving the estimation of $\mu_i$ and $\sigma_i$ in the GEV distribution. There are, however, significant differences between the ML estimates and final model-based posterior estimates of the shape parameter $\xi_i$ and the time trend parameter $\Delta_i$. This is mainly due to shrinkage that is introduced by the unstructured model component in the latent models for $\fat{\phi}$ and $\fat{\gamma}$, and the corresponding PC prior specified for their standard deviations.
While the preliminary site-wise ML estimates of the shape and time trend parameters have very large uncertainties, our fully Bayesian spatial model succeeds in controlling these uncertainties, thus providing significant improvements, i.e., with reduced posterior uncertainty and shrinkage towards more reasonable values. Thus, there is overall an excellent goodness-of-fit with a largely reduced estimation uncertainty compared to site-wise estimation uncertainty.

To investigate the effect of model components on {site-wise} extreme events, e.g., the {marginal} 100-year event, the corresponding quantiles need to be estimated. The quantile function of the GEV distribution has an explicit expression, which makes it straightforward to compute. Inverting \eqref{eq:GEV}, the expression for the $p$-quantile is 
\begin{equation*}
Q(p\mid\mu,\sigma,\xi) = 
\begin{cases} \mu + \sigma[\{-\log(p)\}^{-\xi}-1]/\xi & \xi\neq 0, \\
              \mu - \sigma \log\{-\log(p)\}  & \xi=0,
\end{cases}\qquad p\in (0,1),
\end{equation*}
and the $100$-year event is defined as $Q(0.99\mid\mu,\sigma,\xi)$. In our case, return levels change with time because of the time trend. Therefore, in our context, the 100-year event is the temporally-varying $0.99$-quantile. The multiplication effect of each model component (covariates and spatial effects) on the 100-year event, with all other components being held fixed, is given in Table~\ref{tab:quartileChanges}. 
 \begin{table}[t!]
\rowcolors{2}{gray!6}{white}
\caption{\label{tab:quartileChanges} The effects of the covariates and the spatial model components in the final model on the 100-year event ($0.99$ quantile) of the GEV density. The effect of a given covariate (or a given spatial model component) on the 100-year event is measured by 
comparing the 100-year event computed with the median value of the covariate (or the spatial model component)
to the 100-year event computed with the 1st quartile and 3rd quartile of the covariate (or the spatial model component) while keeping other covariates and random effects fixed. 
The table reports the multiplicative effect on the 100-year event.}
\centering
\begin{tabular}[t]{l|rr}
\hiderowcolors
\toprule
 & \multicolumn{2}{c}{\textbf{$100$-year event}} \\
Covariate & 1st & 3rd \\
\midrule
\showrowcolors
AREA & 0.502 & 2.035  \\
SAAR & 0.747 & 1.390 \\
FARL  & 0.910 & 1.046\\
BFIHOST & 1.071 & 0.905 \\
URBEXT &  1.002 & 0.990\\
FPEXT    &  1.055 & 0.953 \\
Spatial $\psi$ &  0.835 & 1.176\\
Spatial $\tau$ &  0.935 & 1.058\\
\bottomrule
\end{tabular}
\rowcolors{2}{white}{white}
\end{table}
Note that the numbers reported in Table~\ref{tab:quartileChanges} do not change with time. In this table, the parameters are set equal to their posterior means. The unstructured model components, $\fat{\epsilon}_{\psi}$, $\fat{\epsilon}_{\tau}$ and $\fat{\epsilon}_{\phi}$, and the spatial model components, $\fat{u}_{\psi}$ and $\fat{u}_{\tau}$, are set equal to zero when the effects of the covariates are quantified. Table \ref{tab:quartileChanges} shows how the 100-year event changes when covariates or spatial model components vary from their median values to their lower or upper quartiles. As above, AREA appears to be the most important covariate, followed by SAAR.

\begin{figure}[t!]
\centering
\includegraphics[width=0.9\textwidth]{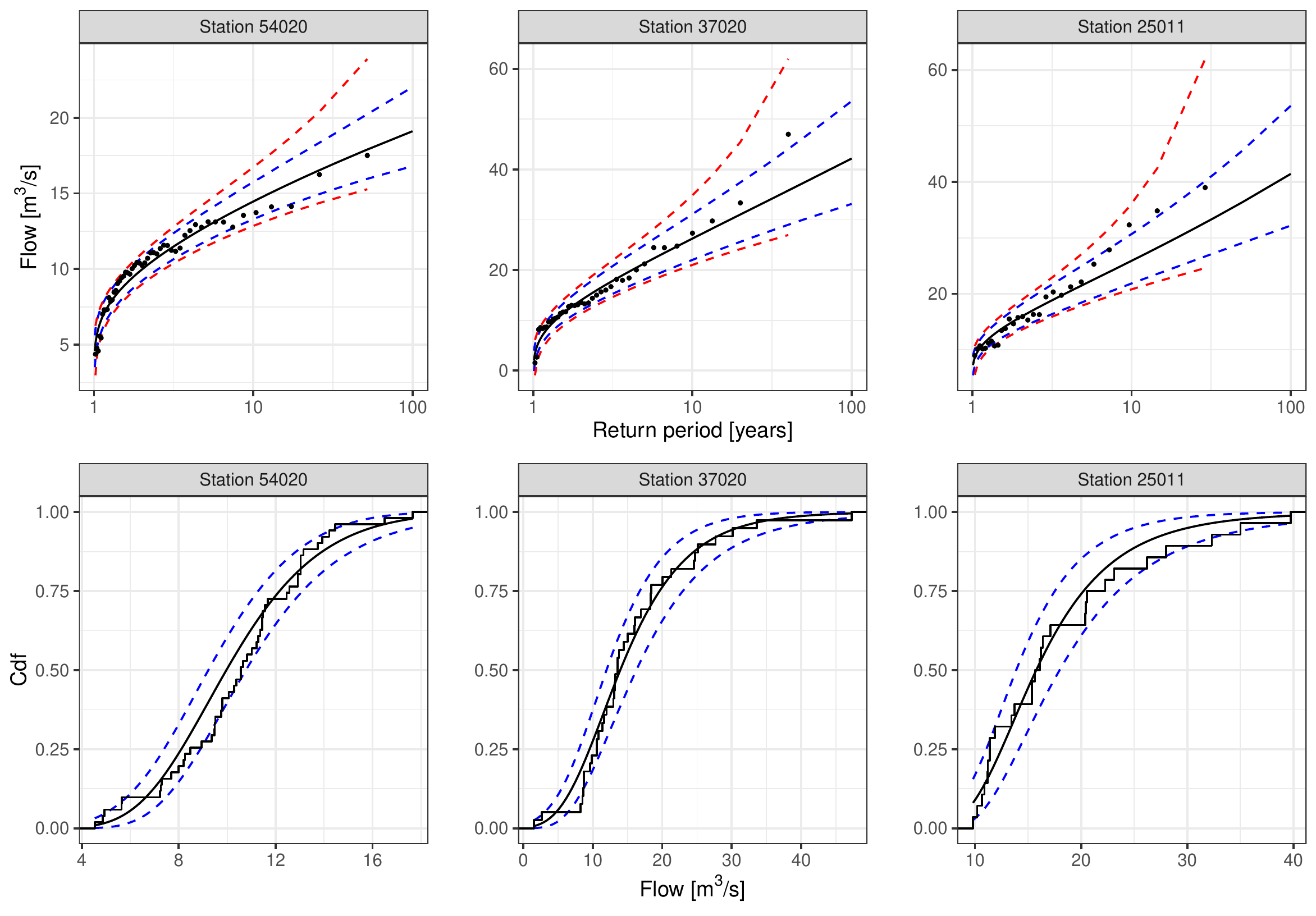}
\caption{Top row: Estimates (i.e., posterior means (solid black line)) of return levels as a function of the return period for three randomly selected stations. The observations at each station were corrected with respect to the posterior estimate of the time trend at the station, i.e., the trend was removed to match a stationary model. The blue dashed lines show 95\% credible intervals for the return levels. The ordered observations (after trend correction) as a function of return period (black points) are shown along with 95\% prediction intervals for the ordered observations (red dashed lines), which represent the variability of the order statistics.
Bottom row: Empirical (black step-wise curves) and fitted (solid black lines) cumulative distribution functions of the annual maximum flow data along with 95\% credible intervals (blue dashed lines) for the fitted cumulative distribution functions. \label{fig:returnPeriodPlots}}
\end{figure}

We then illustrate the estimates of {marginal} return levels and annual maximum flow distributions, under a stationary model, at three stations {(namely, stations 54020, 37020 and 25011)} selected such that the posterior mean of the shape parameter, $\xi$, of the GEV distribution {is approximately $-0.1$, $0.0$ and $0.1$, respectively}. The observations at each site were corrected with respect to the posterior estimate of the time trend at the site, i.e., the time trend component was removed. This was done to facilitate comparison between a stationary model and the data. The top panel of Figure \ref{fig:returnPeriodPlots} shows the posterior means of return levels as a function of the return period, with associated $95\%$ credible intervals, and we also display the order statistics (after time trend correction) for each station with their associated uncertainty (i.e., $95\%$ prediction intervals). 
Notice that the x-axis is expressed on a log-scale. The bottom panel of Figure~\ref{fig:returnPeriodPlots} shows the empirical distribution of the observed annual maxima of the three stations and the posterior mean of the distribution along with $95\%$ credible bands. The relatively good match between the posterior distributions and the empirical distributions indicates that our model fits well overall. Similar results (not shown) were obtained for the other stations. Moreover, by comparing the uncertainty bands in the top panel of Figure~\ref{fig:returnPeriodPlots}, we can see that the uncertainty of our model-based estimates is quite moderate compared to the uncertainty of the (empirical) order statistics. 

\subsection{Predictive performance assessment and model comparison}\label{sec:predictiveperfmodelcomparison}

{To assess the performance of the proposed model and other natural alternative approaches, we design an extensive cross-validation study in which we evaluate the out-of-sample skill of the posterior predictive distributions. To allow evaluation in space and in time, we cross-validate by removing from the training set the data after 2000 at all stations, as well as the six stations depicted in Figure \ref{fig:newfig9} altogether. 
\begin{figure}[t!]
\centering
\includegraphics[width=0.9\textwidth]{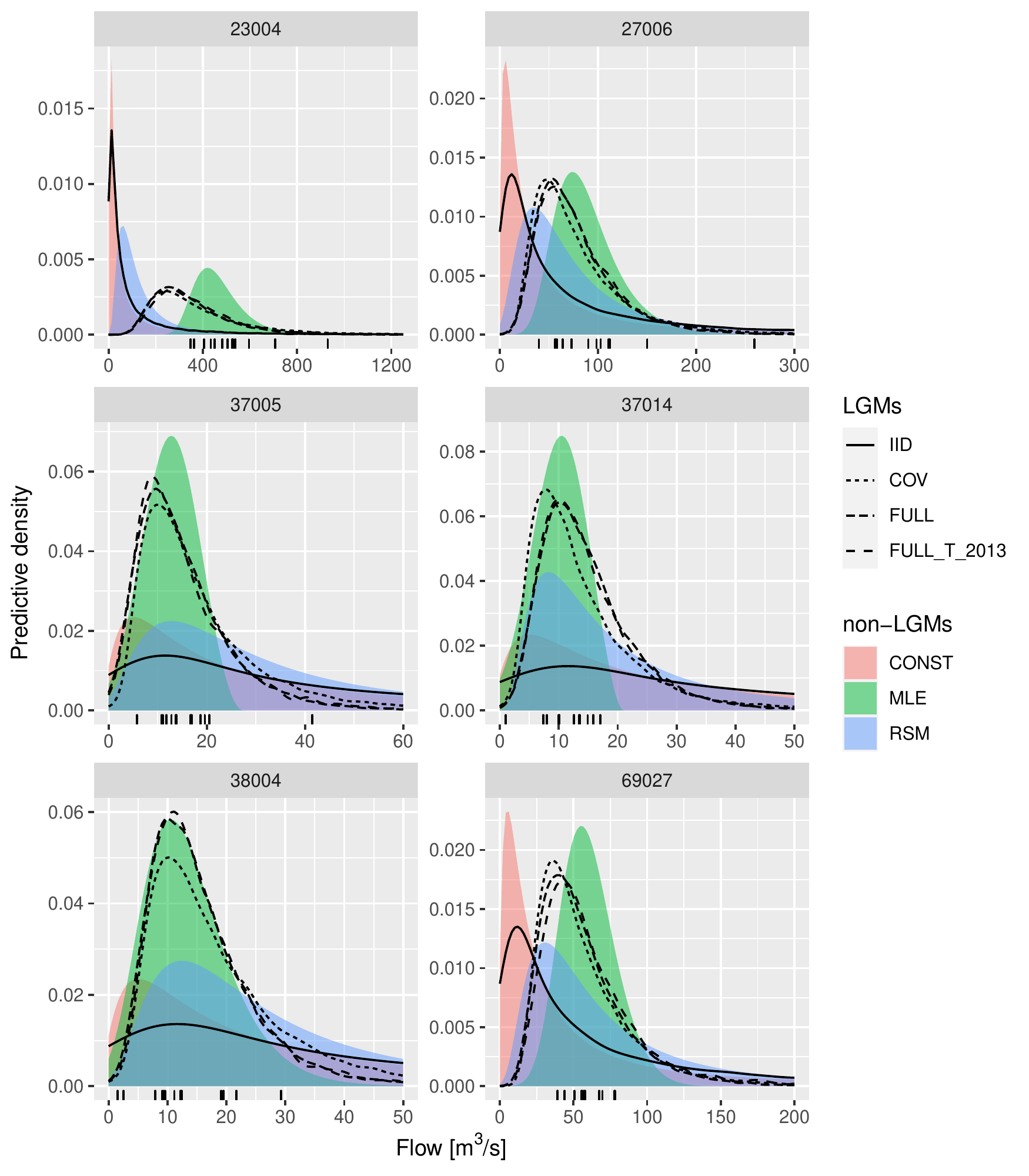}
\caption{{Posterior predictive densities based on four LGMs, the constant model and the response surface model at six stations that where removed from the training data. The long-dashed density (``FULL\_T\_2013'') corresponds to the full LGM with trend, i.e., FULL+T, for the year 2013. The green density corresponds to the ML estimates of the GEV parameters based on site-wise observations and provides a reference (i.e., ``ideal'' prediction) for the out-of-site-out-of-time predictions.}}
\label{fig:newfig9}
\end{figure}
To homogenize the data, we only considered stations that have training data before 1980 and complete test data up to 2013, which leaves 351 stations (compared to 554 in the full dataset). Predictions of the left-out data are evaluated using the logarithmic score \cite[also ``log-score'' or ``ignorance'', see][]{lindley_1985, Roulston_2002}. The log-score is here defined as the negative dual logarithm of the predictive density $p(\cdot)$ assigned to the observed outcome $y$:
\begin{equation} 
{\rm LOGS}(p, y) = -\log_2 p(y) 
\end{equation}
The log-score is negatively oriented, indicating better predictions by smaller
values. We summarize the predictive performance by averaging log-scores over the test data, and compare predictive performance between models by average log-score differences. To give an indication of the robustness of the average log-score differences, we also report their estimated standard errors. To account for correlations in the calculation of standard errors, we replaced the full sample size by an effective sample size $n_{\mathrm{eff}} = n_{\mathrm{eff,time}} \times n_{\mathrm{eff, space}}$. We verified that temporal correlation of log-score differences is low and hence chose $n_{\mathrm{eff,time}} = 13$ (the number of years in the test data). The effective spatial sample size $n_{\mathrm{eff,space}}$ was estimated
as 50, based on spatial variogram ranges of log-score differences.}

{We include the following models in the comparison, all fitted to the same
training data:
\begin{itemize} 
\item CONST: A stationary, spatially invariant and time-constant, GEV distribution fitted to all available
training data by maximum-likelihood estimation.
\item MLE: Individual GEV distributions fitted to training data at each
location separately, using maximum likelihood estimation. This method is not applicable for out-of-site predictions.
\item RSM: A spatial response surface GEV model, using the same covariates as
in our full model (Table \ref{tab:covariatesSelected}) plus polynomials in latitude
and longitude of order 2, 2, 1 for $\mu$, $\sigma$, and $\xi$, respectively, to account for
spatial variability not accounted for by the covariates.
\item IID: An LGM with unstructured spatial random effects $\epsilon$ only included in the GEV model parameters at the latent level.
\item COV: An LGM with unstructured spatial random effects $\epsilon$, and fixed effects of covariates shown in Table \ref{tab:covariatesSelected} (i.e., IID model plus covariates).  
\item FULL: The fully spatial LGM with all covariates and structured/unstructured spatial random effects, but without any time trend (i.e., COV model plus spatial effects).
\item FULL+T: Same as FULL with an additional linear time trend in the location parameter.
\end{itemize}
CONST is the simplest possible null-model, issuing the same prediction at each site in each year, and thus serves a lower benchmark of predictive skill. The response surface model RSM is included to compare our LGM methodology with a more traditional state-of-the-art frequentist methodology for spatial GEV models, which describes GEV parameters in terms of fixed covariate effects \citep{Davison.etal:2012}. The response surface model was fitted by maximum likelihood via the R package \texttt{SpatialExtremes} \citep{spatialextremes}. The four LGM versions tested in this analysis are nested, and increase in complexity from IID to FULL+T. Predictions by all models except FULL+T are constant in time. Some of the predictions produced probabilities smaller than $2^{-50}$, which we treated as numerically equal to zero, and removed the resulting infinitely large log scores from the analysis.}

{For out-of-time cross validation, data at station $i$ up to the year 2000 is
included in the training data, to make post-2000 predictions at station $i$. For out-of-time out-of-site cross validation, no past data from station $i$ is used to make predictions at station $i$. For the time-constant LGMs (IID, COV, FULL) predictive densities were constructed from 32,000 posterior predictive samples by kernel density approximation with a Gaussian kernel and automatic bandwidth selection (using \texttt{R} function \texttt{bw.nrd0}). For the time-varying LGM (FULL+T) we used only 3,200 posterior predictive samples at each station and each year to reconstruct the predictive density in the same way.}

\begin{table}[t!]
\centering
\footnotesize
\caption{Within-site-out-of-time cross validation. Pairwise average log-score differences between models, and estimated standard errors in brackets. Positive entries indicate that column model (e.g., LGM\_FULL) is ``better'' than row model (e.g., CONST).}
\label{tab:logs-within}
\begin{tabular}{r|r|r|r|r|r|r}
\hline
  & MLE & RSM & LGM\_IID & LGM\_COV & LGM\_FULL & LGM\_FULL\_T\\
\hline
CONST & 2.20 (0.07) & 0.70 (0.03) & 2.23 (0.06) & 2.23 (0.06) & 2.23 (0.06) & 2.24 (0.07)\\
\hline
MLE &  & -1.50 (0.07) & 0.03 (0.03) & 0.04 (0.03) & 0.04 (0.03) & 0.04 (0.03)\\
\hline
RSM &  &  & 1.53 (0.06) & 1.53 (0.06) & 1.53 (0.06) & 1.54 (0.07)\\
\hline
LGM\_IID &  &  &  & 0.00 (0.01) & 0.00 (0.01) & 0.01 (0.02)\\
\hline
LGM\_COV &  &  &  &  & -0.00 (0.00) & 0.01 (0.02)\\
\hline
LGM\_FULL &  &  &  &  &  & 0.01 (0.02)\\
\hline
\end{tabular}
\end{table}

{Model CONST was included as a simple null model. All models outperform it significantly. In terms of the within-site-out-of-time predictions, the site-wise MLE estimates predict the observations better than the response surface model, but give slightly worse predictions than the LGMs; see Table \ref{tab:logs-within}. The four LGMs (IID, COV, FULL, FULL+T) are practically the same and all are adequate for within-site predictions, though FULL+T is marginally better. Thus, a Bayesian framework slightly improves over a likelihood-based framework, but covariates and spatial effects do not improve the within-site predictions. We conclude that, in terms of average log-score comparisons, using covariates and spatial effects is unnecessary if site-specific observations are available while the time trend gives a marginal improvement.}

\begin{table}[t!]
\centering
\footnotesize
\caption{Out-of-site-out-of-time cross validation. Note that MLE is within-site, and added here only for reference. See the caption of Table~\ref{tab:logs-within} for further details.}
\label{tab:logs-out}
\begin{tabular}{r|r|r|r|r|r|r}
\hline
  & MLE & RSM & LGM\_IID & LGM\_COV & LGM\_FULL & LGM\_FULL\_T\\
\hline
CONST & 1.97 (0.16) & 0.61 (0.08) & -0.08 (0.05) & 1.48 (0.12) & 1.54 (0.12) & 1.54 (0.13)\\
\hline
MLE &  & -1.36 (0.15) & -2.05 (0.14) & -0.48 (0.09) & -0.43 (0.08) & -0.43 (0.08)\\
\hline
RSM &  &  & -0.69 (0.07) & 0.88 (0.10) & 0.93 (0.10) & 0.93 (0.10)\\
\hline
LGM\_IID &  &  &  & 1.56 (0.10) & 1.62 (0.11) & 1.62 (0.11)\\
\hline
LGM\_COV &  &  &  &  & 0.06 (0.03) & 0.05 (0.03)\\
\hline
LGM\_FULL &  &  &  &  &  & -0.00 (0.01)\\
\hline
\end{tabular}
\end{table}

\begin{figure}[t!]
\centering
\includegraphics[width=\textwidth]{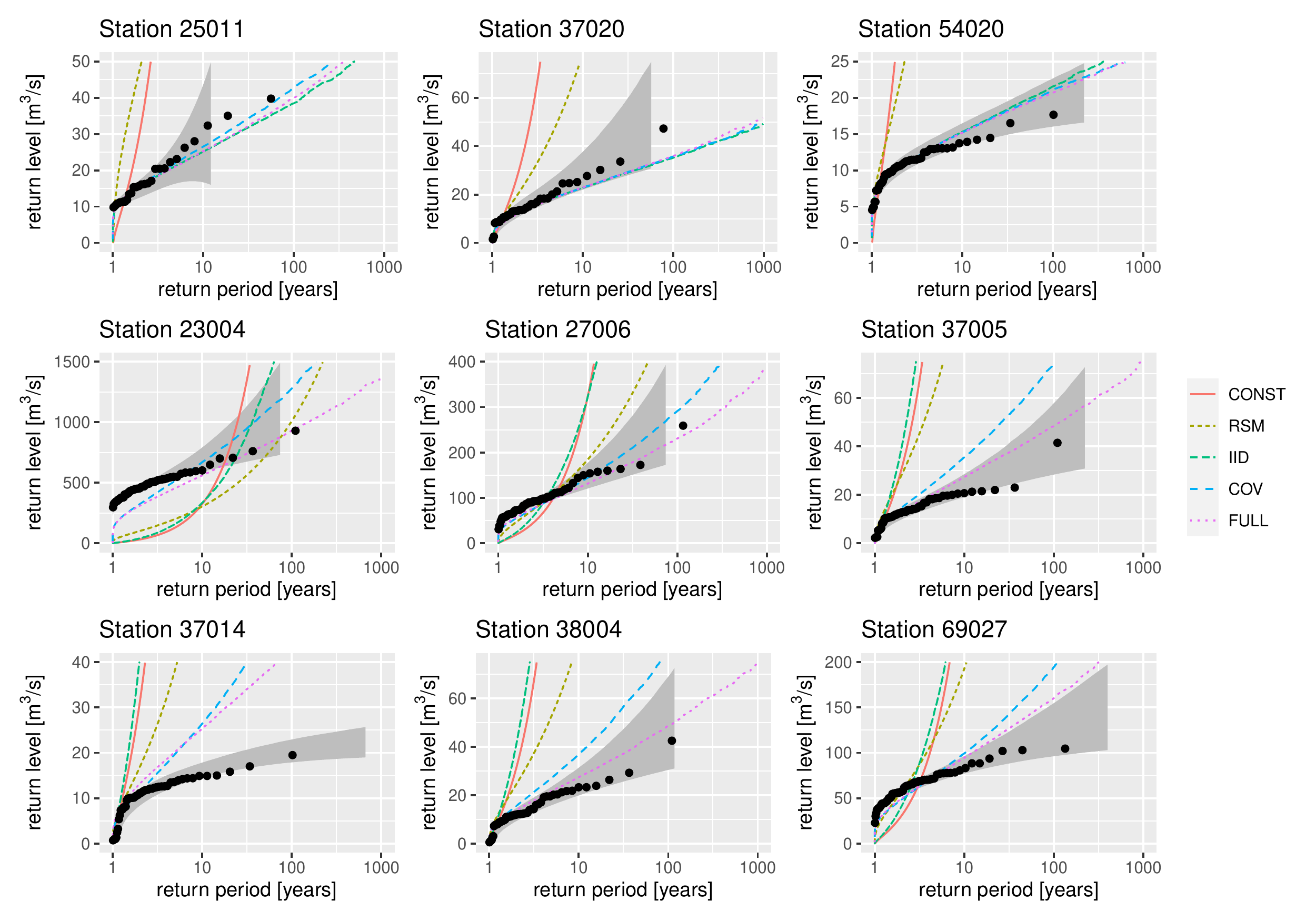}
\caption{{Return level plots based on three of the LGMs (LGM with time trend excluded), the constant model and the response surface model for three within-sample stations and  six stations that where removed from the training data (same as in Figure~\ref{fig:newfig9}). The black dots show the ordered data and the grey bands show 95\% confidence intervals for the return level based on the within-site data.}}
\label{fig:newreturnlevelplot}
\end{figure}

{Out-of-time-out-of-site, the latent Gaussian model IID has similar predictive skill as model CONST, and is outperformed by the other models; see Table~\ref{tab:logs-out}. The site-wise MLEs cannot be produced out-of-site, but were included in the table as an upper reference point of predictive skill that could be achieved if local data were available.  The response surface model RSM outperforms the IID model because of the added information from local covariates that cannot be captured out-of-site by an independent random effect. Similar to the within-site cross validation, all of COV, FULL, and FULL+T outperform RSM. 
The relatively small improvement of including spatial random effects (FULL versus COV) 
might be due to the relatively large number of covariates, whose spatial
structures are enough to account for the spatial variation in the model parameters.
The FULL+T model is on a par with the FULL model, so there is here no noticeable
additional benefit of including a linear trend.}

{Figures~\ref{fig:newfig9} and \ref{fig:newreturnlevelplot} provide a
visual confirmation that the full LGM (with or without the time trend component) does slightly better than the other models in terms of out-of-site predictions, as its densities and return level curves are always closest to the densities and return level curves stemming from the data. These results are in line with the out-of-site log-score results presented in Table~\ref{tab:logs-out}. Figure~\ref{fig:newreturnlevelplot} shows that the three LGMs without time trend perform equally well in terms of within-site predictions, while the constant model and the response surface model perform very poorly, matching the results in Table~\ref{tab:logs-within}.  
}

{We now further discuss the out-of-site predictions for Station 23004 presented in Figure~\ref{fig:newfig9}. In this case the predictors, i.e., the covariates and the spatial components, of the three most complex LGMs are such that the $\epsilon$ term for the log-location parameter $\psi$ is positive and around $0.2$ in size, i.e., not particularly large values since the standard deviation of $\epsilon_\psi$ is estimated as $0.247$. However, since the $\epsilon$ term is unknown, this implies that the out-of-sample posterior predictive density for this site will be centered at a level about $20$\% smaller than the center of the data density. However, the variability in the data density at the site is smaller than the variability in the posterior predictive density due to the four unknown $\epsilon$ terms, namely, $\epsilon_\psi$, $\epsilon_\tau$, $\epsilon_\phi$, and $\epsilon_\gamma$, and thus, most of the observations fall under the posterior predictive density, mainly its upper tail since $\epsilon_\psi$ was positive. The LGM with unstructured terms only does poorly here. This is due to the standard deviation of the $\epsilon$ being much larger than in the LGMs with covariates and/or spatial components. The response surface model does not perform as well as the three most complex LGMs but does better than the LGM with unstructured terms only, underlining the importance of covariates.}

\subsection{Time trend interpretation and assessment}
\begin{figure}[t!]
\centering
\includegraphics[width=0.75\textwidth]{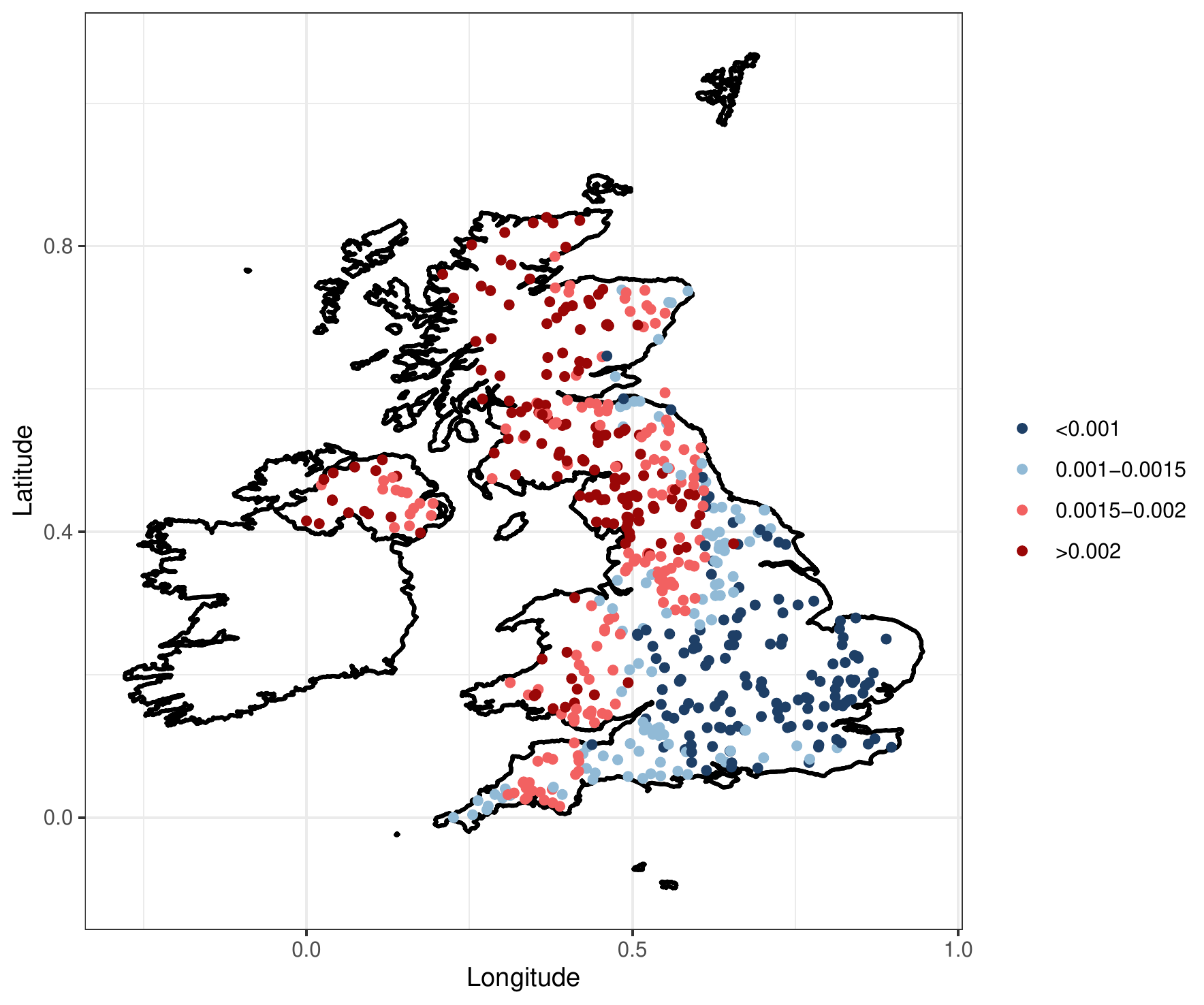}
\caption{{Posterior estimates of $\Delta_i$ across the UK, presented with a $4$-color scheme that splits trend values at  $\Delta_i=0.010,0.015,0.0020$.}}
\label{fig:trendspatial}
\end{figure}

{Figure \ref{fig:trendspatial} shows the spatial structure of the time trend parameter $\Delta$ based on its posterior estimates.
The trend estimates correspond to a $0.1$\% to $2.8$\% increase per decade with a median value of around $1.5$\% per decade. In Figure \ref{fig:trendspatial} it can be seen that the trend is greater in the northern and southwestern part of the UK, while smaller trends are observed in the southeastern part. This is in line with the results in \citet{bloschl2019}, who also found greater trends in the northern part of the UK and smaller trends in the southeastern part. However, their results for the southwest part and the part below the middle of the UK are somewhat different from ours. 
The analysis of \citet{bloschl2019} indeed resulted in negative trend values in the middle part of the UK, while here our trend estimates are always positive. Due to the latent Gaussian structure of our proposed model, in particular the part involving $\Delta$, the values of the corresponding posterior estimates spread less than those found in \citet{bloschl2019}, the estimates above $0.0015$ being usually smaller than those in \citet{bloschl2019} and none of the estimates below $0.0015$ being negative. We believe the main reason for this disagreement is that \citet{bloschl2019} performed a site-by-site analysis (with independent GEV fits at each site), while our more complex latent Gaussian model incorporates informative covariates and spatial effects at the latent level, which allows a drastic reduction in parameter uncertainty through shrinkage by borrowing strength across neighboring sites. This is especially helpful at sites with small sample sizes and can prove critical to avoid estimating unrealistic trend values.}

{In \citet{dadson2017} various aspects of flood management in the UK were considered. They point out that: (i) winter precipitation has been increasing during the period 1950--2015 and the same is true for the summer precipitation for the period 1975--2015 \citep{ALEXANDER2000142}; and (ii) the annual mean flood index has been increasing during the period 1990--2015 \citep{wilby2013reconstructing}.  \citet{bloschl2019} show that an upward trend over the period 1960--2010 in a flood index and in maximum 7-day precipitation in the UK. Generally speaking, our results are therefore in line with the published literature, which provides further evidence that our proposed LGM is indeed able to detect the time trend appropriately}

\subsection{Spatial dependency at the data level and estimation of return levels for spatial aggregates}\label{sec:spatialdependency}

{The assumption of independence at the data level is problematic when spatially
extended quantities such as aggregates or co-exceedances are of interest.  To
reliably predict, say, the annual maximum flow summed over a few neighboring
catchments, we have to take into account not only their individual marginal
distributions, but also their spatial dependency. Working under a modeling
framework that ignores spatial dependence at the data level will inevitably provide biased predictions for spatially aggregated quantities.}

{Accounting for spatial dependence at the data level in a fully Bayesian way
leads to a more complicated inference framework, which is outside the scope of
this paper. However, we can predict annual maximum flow jointly at multiple
stations in a way that respects the spatial dependence structure by reordering independently
drawn posterior predictive samples based on the empirical copula obtained from
historical observations; see \citet{Clark_2004} and \citet{Schefzik_2013} for related approaches. Our proposed method starts
from a set of $M$ historical observations available at each of $L$ stations of
interest. The rank order of the observations at each site is calculated. Then, $M$ independent samples are drawn from each of the $L$ univariate posterior predictive distributions of the GEV model fitted to each site. These $M$
independent samples are subsequently reordered to have the same
rank order as the $M$ historical observations. The reordered samples have the
same pairwise Spearman rank correlations as the historical observations, and
thus mimic the observed spatial dependency. By repeatedly drawing $M$
independent samples at each site and rank-reordering them in the same way, the
posterior predictive sample size can be increased indefinitely.}

{For illustration we have selected 15 nearby stations in the Eastern part of the UK and
considered the sum of their annual flow maxima as a prediction target. The
aggregate of annual maxima over several catchments might be of interest to
insurers or infrastructure planners as an indicator of compound regional flood
risk. After fitting the full LGM (that assumes conditional independence at the data level),
we generated independent posterior predictive samples at each location. We
estimated the predictive distribution of the spatially aggregated annual
maximum flow, based on conditionally independent samples, and based on samples that were
rank-reordered to restore spatial correlation as outlined above.}

\begin{figure}[t!]
\centering
\includegraphics[width=0.64\textwidth]{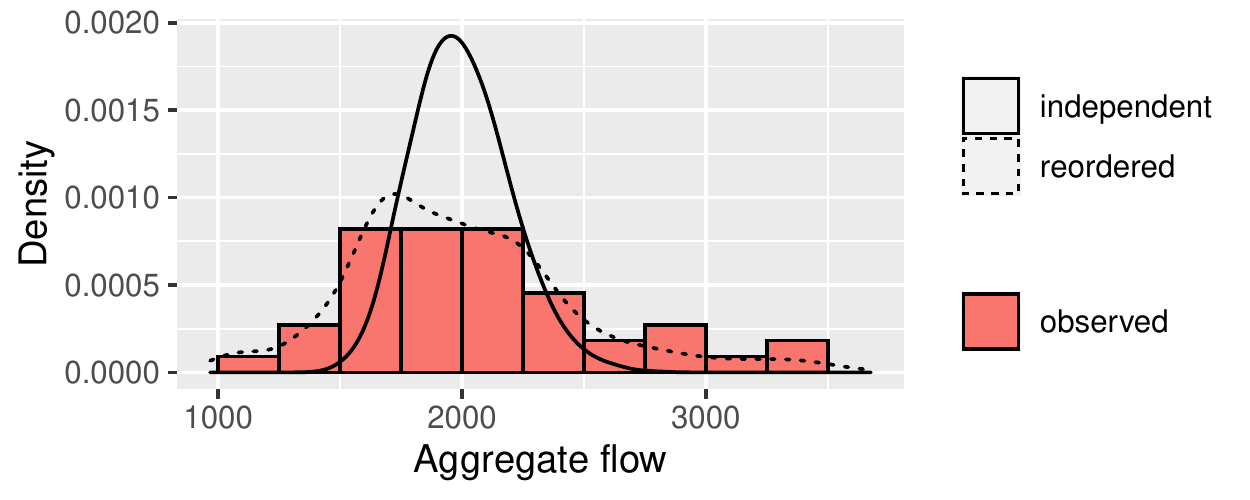}
\includegraphics[width=0.64\textwidth]{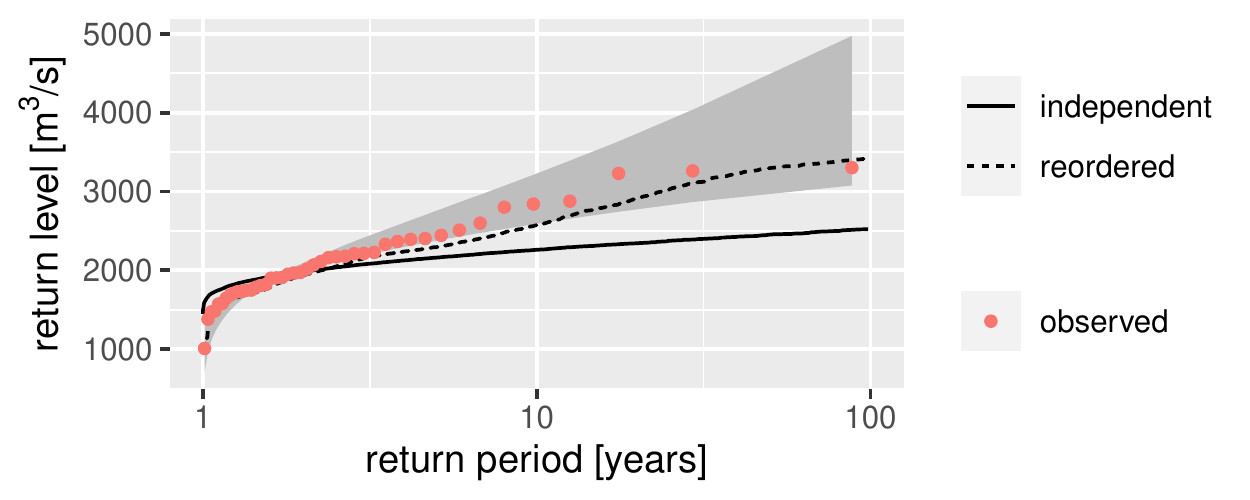}
\caption{{Top: Predictions of aggregate annual maximum flow taken over 15 nearby stations, based on conditionally independent samples and rank-reordered samples. After restoring the spatial dependence in posterior predictive samples, the predictive distribution matches the frequency histogram. Bottom: Return level plots corresponding to the densities in the top panel show that observed return levels are truthfully represented after rank-reordering.}}
\label{fig:rankshuffle}
\end{figure}

{The histogram of observed aggregate maximum flow, and density estimates of
predictive samples before and after rank-reordering are shown in Figure~\ref{fig:rankshuffle} (top panel). The posterior predictive distribution of
aggregated flow estimated from conditionally independent samples clearly does not reflect the
observed variability.  Both small values and large values are underrepresented compared to the empirical histogram. After rank-reordering, the predictive
distribution has the appropriate width, and correctly reflects the observed
variability of aggregate flow maxima. As a result, return levels of aggregate
maximum flow calculated from the rank-reordered samples are also much more accurate than the return levels calculated
from spatially independent samples (see the bottom panel of Figure~\ref{fig:rankshuffle}).}

{These results suggest that post-processing independent samples by
rank-reordering is a suitable method to restore the previously ignored
dependency at the data level. However, it is worth noting that by ignoring that
dependency during parameter inference, the posterior distributions of the
latent parameters are likely too narrow, and thus overconfident. Moreover,
rank-reordering can only be applied at locations where historical data are
available.  If past data is not available, the rank order has to be estimated
either by interpolating between nearby observed stations, or from a suitable surrogate, such as numerical model simulations.}

\section{Discussion}
\label{ch:discussion}

In this paper, we developed an extended latent Gaussian model based on the generalized extreme-value distribution, designed for {analysis of spatio-temporal flood frequency data}, and we illustrated our approach by application to a large dataset of maximum annual peak flow time series from the UK.
Because of the large dimensionality of the data and the high model complexity, GMRF priors were assumed for latent spatial effects. 
We used Max-and-Smooth \citep{hrafnkelsson2020approx}
for fast posterior inference, and thoroughly verified its accuracy when using a GEV distribution at the data level. This inference approach relies on a Gaussian approximation of the likelihood function, which simplifies inference and provides significant speed-up. 
The covariates for the final model were selected with a simplified model setup that made it possible to use INLA, which led to further substantial reductions in computation time. The selected covariates were all highly significant in the final model, i.e., their $95\%$ posterior intervals did not include zero. 

How to {structure and parameterize} the GEV distribution within a Bayesian hierarchical model is an important and challenging methodological question. 
In this paper a novel multivariate link function for the three parameters of the GEV distribution was proposed, along with a temporal trend parameter included in the GEV location parameter.
A standard logarithmic transformation was used for the location $\mu$, and the confounding between the location $\mu$ and scale $\sigma$ was dealt with by transforming them jointly.
We also proposed a novel transformation for $\xi$ that was constructed based on four criteria that we believe are reasonable when inferring the tail behavior in a wide range of environmental applications. Moreover, the new transformation stabilized the inference of $\xi$, and through an additional beta prior density for each $\xi_i$, its inference can be tailored to each specific application. The time trend was bounded between $-\delta_0$ and $\delta_0$,
with $\delta_0=0.008$, which is physically justifiable and {seemed} reasonable in our context. The posterior estimates of the time trend parameters correspond to an average of $1.5$\% increase per decade in extreme river flow in the United Kingdom, likely due to climate change. {Our results show a spatial structure in the time trend that is similar to that of \citet{bloschl2019}, however, our estimates are usually smaller and all positive. While prior specification and reparametrizations are sometimes regarded as a minor aspect in the entire modeling pipeline, we rather consider it as a crucial element in Bayesian modeling that requires in-depth reasoning and justifiable arguments. The novel transformations and priors proposed in our paper rely on valid physical and statistical considerations, and our paper is a call for a more careful modeling of key parameters (such as the shape $\xi$ and the trend $\Delta$).}

The proposed model was inferred with a Bayesian approach and the data were not transformed before inferring the parameters and thus their uncertainty was properly quantified. Normalization is required for some commonly used flood frequency models such as the index flood method and as a consequence uncertainty is not properly propagated. Under the extended LGM framework used here, it was straightforward to include covariates and additional random effects at the latent level and keep track of the uncertainty of the model parameters.
PC priors were defined for the hyperparameters of the model. The hyperparameters in the extended LGM setup govern the latent parameters which are a part of an additive regression model, and thus the penalization of increased complexity becomes essential to regularize the random effects at the latent level.

The results showed that all parameters of the model converged quickly and most parameters were
well defined. In particular, all hyperparameters had small posterior standard deviations with respect to their posterior means. The results showed that the spatial model components for $\fat{\psi}$ and $\fat{\tau}$ explained more than half of the otherwise unexplained variability, which indicates that they are essential in our model. 
The return period of extreme floods (under a stationarity assumption) was investigated by computing quantiles of the fitted GEV distribution and taking their uncertainty into account. Posterior predictive distributions were computed for six gauging stations and our results showed that our proposed model is useful for predicting extreme flow data within ungauged catchments. 

Our analysis demonstrates that the LGM framework is very flexible and powerful for predicting extreme flow data within gauged and ungauged catchments, by efficiently borrowing strength across locations. However, the underlying conditional independence assumption of the data with respect to latent parameters makes it unsuitable when dependence across catchments is present and predictions on the joint behavior of the catchments are needed. \citet{Sang2010} suggested relaxing the conditional independence assumption by using a Gaussian dependence structure at the data level, but this approach does not properly capture extremal dependence. For this purpose, more complex and specialized extreme-value models are required. In the case of strong extremal dependence, max-stable processes \citep{Padoan.etal:2010,Davison.etal:2012,Huser2014,asadi2015extremes,Vettori.etal:2019,Davison.etal:2019} have proven to be useful for modeling spatial block maxima, while in the case of weakening extremal dependence, broader classes of max-infinitely divisible processes \citep{Huser.Wadsworth:2020,Huser.etal:2021,Bopp.etal:2021} have recently been proposed. These models, however, are usually much more intensive to fit than our proposed extended LGM and further research is needed to make them applicable in large dimensions as in this paper. Nevertheless, we computed the effect of neglecting dependence at the data level in a simplified setting (see {Section~4 of} the Supplementary Material) and found that this effect was mostly moderate in our application. These results might also be exploited to correct the uncertainty estimates provided by the conditional independence model. {Furthermore, we here also developed a novel copula-based post-processing approach to correct the dependency in posterior predictive samples and thereby accurately estimate return levels of spatial aggregates. Our proposed approach, however, relies on computing the empirical copula of historical data at the target locations, so future research is needed to extend it for the estimation of spatial return levels involving unobserved locations.}

We believe that our proposed latent Gaussian model and its inference scheme have features that are important for flood frequency analysis in general. In particular, predictions of extreme events in the case of both gauged and ungauged catchments are improved due to (i) reducing the uncertainty in the shape and time trend parameters by transforming them and regularizing the transformed parameters with unstructured model components; (ii) modeling the location and scale parameters spatially and utilizing the correlation between these two parameters; (iii) applying the fast approximated inference scheme, making it possible to evaluate large number of potential regression models and to perform extensive cross-validation to achieve good prediction properties. In our view, future research efforts in flood frequency analysis should focus on the development of fast inference schemes for models that take extremal dependence into account. 



\bibliographystyle{CUP} 
\bibliography{bibTex}

\end{document}